\documentclass[aps, prx, twocolumn, superscriptaddress, floatfix,english]{revtex4-2}
\usepackage{graphicx}
\usepackage{amsmath}
\usepackage{tikz}
\usepackage{braket}
\usepackage{pgfplots}
\usepackage[version=4]{mhchem}
\pgfplotsset{compat=1.17}
\usepackage{subcaption}
\usetikzlibrary{shapes.geometric}
\usetikzlibrary{shadows,patterns,perspective}
\usetikzlibrary {decorations,decorations.text}
\usetikzlibrary{decorations.pathreplacing}
\usepackage{float}
\usepackage{hyperref}
\usepackage{xcolor}

\definecolor{linkcolor}{RGB}{0, 0, 255}      % Blue color for links
\definecolor{citecolor}{RGB}{0, 128, 0}     % Green color for citations
\definecolor{urlcolor}{RGB}{255, 0, 0}       % Red color for URLs

\hypersetup{
    colorlinks=true,
    linkcolor=linkcolor,
    citecolor=citecolor,
    urlcolor=urlcolor,
    linktoc=all,  % Set links in the table of contents to be colored as well
    pdfborder={0 0 0}  % Removes the box around links
}

\DeclareCaptionJustification{justified}{\leftskip=0pt \rightskip=0pt \parfillskip=0pt plus 1fil}

\begin{document}
\definecolor{dy}{rgb}{0.9,0.9,0.4}
\definecolor{dr}{rgb}{0.95,0.65,0.55}
\definecolor{db}{rgb}{0.5,0.8,0.9}
\definecolor{dg}{rgb}{0.2,0.9,0.6}
\definecolor{BrickRed}{rgb}{0.8,0.3,0.3}
\definecolor{Navy}{rgb}{0.2,0.2,0.6}
\definecolor{DarkGreen}{rgb}{0.1,0.4,0.1}
\title{Kondo effect in
 the isotropic Heisenberg spin chain}

\author{Pradip Kattel}
\thanks{These authors contributed equally.}
\affiliation{Department of Physics, Center for Material Theory, Rutgers University,
Piscataway, NJ 08854, United States of America}
\email{pradip.kattel@rutgers.edu}

\author{Parameshwar R. Pasnoori}
\thanks{These authors contributed equally.}
\affiliation{Department of Physics, University of Maryland, College Park, MD 20742, United
States of America}
\affiliation{Laboratory for Physical Sciences, 8050 Greenmead Dr, College Park, MD 20740,
United States of America}
\email{pparmesh@umd.edu}

\author{J. H. Pixley}
\affiliation{Department of Physics, Center for Material Theory, Rutgers University,
Piscataway, NJ 08854, United States of America}

\author{Patrick Azaria}
\affiliation{Laboratoire de Physique Th\'orique de la Mati\`ere Condens\'ee, Sorbonne Universit\'e and CNRS, 4 Place Jussieu, 75252 Paris, France}
\date{\today}

\author{Natan Andrei}
\affiliation{Department of Physics, Center for Material Theory, Rutgers University,
Piscataway, NJ 08854, United States of America}

\begin{abstract}
We investigate the boundary effects that arise when spin-$\frac{1}{2}$ impurities interact with the edges of the antiferromagnetic spin-$\frac{1}{2}$ Heisenberg chain through spin exchange interactions. We consider both cases when the couplings are ferromagnetic or anti-ferromagnetic. We find that in the case of antiferromagnetic interaction, when the impurity coupling strength is much weaker than that in the bulk, the impurity is screened in the ground state via the Kondo effect. The Kondo phase is characterized by the Lorentzian density of states and a dynamical scale, the Kondo temperature $T_K$, is generated. As the impurity coupling strength increases, $T_K$ increases until it reaches its maximum value $T_0=2\pi J$ which is  the maximum energy carried by a single spinon. When the impurity coupling strength is increased further, we enter another phase, the bound mode phase, where the impurity is screened in the ground state by a single particle bound mode exponentially localized at the edge to which the impurity is coupled. We find that, in contrast to the Kondo phase, the impurity can be unscreened by removing the bound mode. This costs an energy $E_b$ that is greater than $T_0$. There exists a boundary eigenstate phase transition between the Kondo and the bound-mode phases, a transition which is characterized by the change in the number of towers of the Hilbert space. The transition also manifests itself in local thermodynamic quantities - local impurity density of states and the local impurity magnetization in the ground state. When the impurity coupling is ferromagnetic, the impurity is unscreened in the ground state; however, when the absolute value of the ratio of the impurity and bulk coupling strengths is greater than $\frac{4}{5}$, the impurity can be screened by adding a bound mode that costs energy greater than $T_0$. When two impurities are considered, the phases exhibited by each impurity remain unchanged in the thermodynamic limit, but nevertheless the system exhibits a rich phase diagram.

\end{abstract}

\maketitle

\section{Introduction}
\label{sec:intro}

The Kondo effect is a paradigmatic example of a strongly correlated phenomenon. The conventional Kondo system consists of a single localized spin impurity placed in a metal \cite{hewson1997kondo}.  It interacts with the conduction electrons, modeled as a free bath of electrons,  via an antiferromagnetic spin-exchange coupling whose strength depends on the energy scale being observed. In particular, the coupling strength increases as the energy scale is decreased,
thereby leading to a non-perturbative ground state \cite{kondo2012physics}.

 The  description of the electrons in a metal as a non-interacting  gas, a Fermi liquid,  is valid in a three dimensional metal, as the interactions among them lead typically only to a renormalization of their parameters \cite{landau1957theory}. In a one-dimensional metal, however, any interaction leads to signatures of non-Fermi-liquid behavior. These non-Fermi liquid behaviors are captured by the Luttinger liquid theory \cite{haldane1981luttinger}. The quest to understand the effect of a Kondo impurity in a Luttinger liquid has a long history \cite{furusaki1994kondo,lee1992kondo,schiller1995exact,frojdh1995kondo,fendley1996unified}. Using a combination of techniques ranging from  Abelian
bosonization \cite{lee1992kondo}, perturbative renormalization group based approach \cite{furusaki1994kondo}, and  boundary CFT \cite{frojdh1995kondo},  the problem has been studied in  continuum models with the impurity placed in the bulk. However, the problem is yet to be resolved because the boundary CFT predicts two possible strong-coupling fixed points: the Fermi liquid universality class or non-Fermi liquid behavior, and it is not clear whether the extrapolation
of the perturbative scaling equations into the strong coupling done using Poor man's scaling is justified.

Independently, it was realized that one version of the problem, spin chains with magnetic impurities, can be solved exactly via \textit{Bethe Ansatz} \cite{andrei1984heisenberg,eckle1997absence,Schlottmann,sacramento1993thermodynamics}. These models capture some aspects of the problem since the low-energy behavior of spin chains is described by a Luttinger liquid \cite{bosonizationLiu}, which corresponds to the spin component of a one-dimensional spin-charge decoupled gas of electrons. In another version of the problem, when impurities are at the boundary, the model captures some aspect of Kondo physics in interacting media \cite{eckle1997absence,laflorencie2008kondo}. 
In another approach, Wang studied the finite-length spin-$\frac{1}{2}$ Heisenberg model with a spin-$S$ magnetic impurity coupled to each edge of the chain with equal coupling using \textit{Bethe Ansatz} in \cite{wang1997exact}. 
A modified version of the problem was  studied in \cite{frahm1997open}.  Moreover, various numerical techniques have been employed to investigate the effect of magnetic impurities on the spin chain \cite{XXZkondoDMRG,numFulde95,numfulde97,XXZKondogaplessfurusaki,Eggert2001num}. Later, similar techniques were applied to study quantum impurities in various interacting media \cite{rylands2020exact,pasnoori2020kondo,rylands2016quantum,rylands2017quantum,rylands2018quantum,oz2019hubbard,links1999integrability,fan2001integrable,frahm2007integrable}.

 In this paper, we solve the antiferromagnetic spin-$\frac{1}{2}$ Heisenberg chain coupled to spin-$\frac{1}{2}$ impurities at the edges. Using Bethe Ansatz and DMRG we address several issues left open in previous studies. In particular, having the model defined on the lattice, rather than in the continuum, gives us access to the problem at all energy scales. In particular, we find that not all phases allow description by CFT.

 The Heisenberg model describes many physical compounds such as \ce{SrCo2V2O8}, \ce{KCuF3}, \ce{CuCl2*2NC5H5}, \ce{Cu(NH3)4SO4*H2O} \textit{etc.} \cite{scheie2022quantum,wang2018experimental,steiner1976theoretical,nagler1991spin}. As experimental samples are likely to have defects and impurities, it is important to understand the roles of defects in these systems. The effect of impurities has been experimentally studied in a few quasi one-dimensional systems \cite{chakhalian2004impurity,aczel2007impurity}.

We first consider the case of one impurity coupled to an edge of the chain, and investigate the system using Bethe Ansatz and DMRG. Subsequently, we shall consider the system with two impurities and analyze it using the Bethe Ansatz. The one-impurity model is described by the Hamiltonian 

\begin{equation}
    H=\sum_{j=1}^{N-1}  J \vec{\sigma_j} \cdot \vec{ \sigma}_{j+1}+ J_{imp}\vec\sigma_1\cdot\vec \sigma_L.
    \label{xxximp}
\end{equation}

When the impurity coupling is antiferromagnetic, we find there are two phases depending on the relative values of the impurity and bulk coupling strengths. When the ratio of the impurity coupling strength $J_{\mathrm{imp}}$ and the bulk coupling strength $J$ is sufficiently smaller than $1$, the system exhibits a genuine multiparticle Kondo effect characterized by the ratio of impurity and bulk density of state (DOS) taking a Lorentzian-like form that reached a peak at $E = 0$. When the ratio between the impurity and the bulk coupling strengths $(J_{\mathrm{imp}}/J)$ is increased such that it approaches $1$, the impurity DOS loses the Lorentzian form and takes the form of a constant function. This is to be expected because when $(J_{\mathrm{imp}}/J) = 1$, the impurity no longer exists, as it becomes a part of the chain, increasing the number of sites of the chain by one. As the ratio $(J_{\mathrm{imp}}/J)$ increases further, so it lies between $1$ and $4/3$, the impurity DOS  peaks at $T_0=2\pi J$, which corresponds to the maximum energy of a single spinon. As the ratio $(J_{\mathrm{imp}}/J)$ approaches $4/3$, the DOS of the impurities takes the form of a delta function centered at $E_b$ indicating that the screening of the impurity is due to a single mode. For $0<(J_{\mathrm{imp}}/J)<4/3$, the many-body screening described above is characterized by the renormalized Kondo temperature $T_K<T_0$.  As the ratio $(J_{\mathrm{imp}}/J)$ is increased past $4/3$, the impurity remains screened in the ground state due to a single mode bound to the impurity, which can now be removed at  energy cost $E_b>T_0$. We refer to this phase as the `Antiferromagnetic Bound mode (ABM) phase'. This phase cannot be seen in the low-energy continuum description of the model \cite{laflorencie2008kondo}, occurs when the impurity energy scale is beyond the cutoff scale imposed by the lattice spacing, and hence it is characterized by the presence of a high-energy mode exponentially localized near the impurity. In the case of ferromagnetic coupling, the impurity is unscreened in the ground state for all values of $(J_{\mathrm{imp}}/J)$. For $|J_{\mathrm{imp}}/J|>4/5$, there exists a high energy mode exponentially localized near the impurity which screens the impurity in the excited states, and hence we name this phase the `Ferromagnetic Bound mode (FBM) phase'. The phase corresponding to $(J_{\mathrm{imp}}<J)$, where the impurity cannot be screened is named the `Unscreened Phase (US)'.

Also the 
nature of the excited states varies from phase to phase. In the Kondo and the unscreened phases, only bulk excitations are present, while in the bound-mode phases both boundary and bulk excitations are present. In particular, each of the bound-mode phases displays two separate towers of excited states, one where all the states contain an unscreened impurity and the other tower  where all states contain an impurity screened by the local bound mode. These towers of excited states were also observed in the XXX chain
with boundary magnetic field studied in \cite{pasnoori2021boundary}.

Turning to consider the case where each edge of the chain is coupled to an impurity, the system is described by the Hamiltonian
\begin{equation}
    H=J\sum_{j=1}^{N-1}\vec\sigma_j\cdot\vec\sigma_{j+1}+J_L\vec\sigma_1\cdot\vec\sigma_L+J_R\Vec\sigma_N\cdot\vec\sigma_R.
    \label{ham2imp}
\end{equation}

\begin{figure}
    \centering
    \begin{tikzpicture}[scale=0.55, every node/.style={scale=0.65}]
 %boxes with fading colors
  \fill[left color=dr, right color = dy!20] (0,0) rectangle (5,10);
  \fill[left color=dy!30, right color = dy] (5,0) rectangle (7,10);
  \fill[left color=dg, right color=db!50] (7,0) rectangle (9,10);
  \fill[left color=db!30, right color=db] (9,0) rectangle (14,10);

  %axes and axes labels
\node at (-1,2.5) {\footnotesize{Energy}};
\node at (3.25,-1) {\footnotesize{$\frac{J_\mathrm{imp}}{J}$}};

\draw [<->, rounded corners, thick, purple] (-1,2.3) -- (-1,-1) -- (2.8,-1);

\draw [<-] (.5,-.5)--(2,-.5);
\draw [->] (12.25,-.5)--(13.75,-.5);

%xticks
\node at (9,-.5) {$\frac{4}{3}$};
\node at (7,-.5) {$0$};
\node at (5,-.5) {$-\frac{4}{5}$};
 \node at (0,-.5) {$-\infty$};
 \node at (14,-.5) {$\infty$};

 \draw[dotted, thick](7,0)--(7,10);
 \draw [dotted,red,thick] (7,4)--(9,4);
\draw [dotted,blue,thick] (5,6)--(7,6);

  \coordinate (A) at (9,4);
  \coordinate (B) at (14,0);
  
  \draw[red,thick] (A) .. controls (12.5,4) and (13.5,3.5) .. (B);

  \coordinate (C) at (5,6);
  \coordinate (D) at (0,10);
  
  \draw[blue, thick] (C) .. controls (1.5,6) and (0.5,6.5) .. (D);

  %adding texts
  \node at (8,9) {Kondo};
  \node at (8,8.5) {Phase};
  \node at (8,7.75) {\footnotesize{Multiparticle}};
  \node at (8,7.4) {\footnotesize{screening of}};
  \node at (8,7.05) {\footnotesize{impurity spin}};

   \node at (6,9) {Unscreened};
  \node at (6,8.5) {Phase};

\node at (11.25,5.75) {Unscreened impurity};
\node at (11.5,5.35) {No boundary string $S^z=\pm \frac{1}{2}$};

    \node at (11.5,9) {Bound mode};
  \node at (11.5,8.5) {Phase};
  \node at (11.5,7.75)
{\footnotesize{single particle bound mode}};
  \node at (11.5,7.40) {\footnotesize{screens the impurity spin}};

   \node at (2.5,9) {Bound mode};
  \node at (2.5,8.5) {Phase};

\draw[thick,blue](9,5)--(14,5);
\draw[thick,red](0,5)--(5,5);

   \node at (2.75,5.75) {Unscreened impurity};
\node at (2.5,5.35) {No boundary string $S^z=\pm \frac{1}{2}$};

\node at (7,3.25) {No boundary excitations};

    \node at (10.5,2) { Ground state is unique with $S_z=0$ };
  
 \node at (2.5,3.3) 
  {\footnotesize{Bound mode screening the}};
  \node at (2.5,2.9) {\footnotesize{ impurity spin occurs}};
  \node at (2.5,2.5) {\footnotesize{ only at high energies}};

  \node at (7.2,5) {0};
   \node at (6.5,4) {$-\pi J$};
  \node at (7.3,6) {${\pi J}$};

 \draw [decorate,decoration={brace,amplitude=10pt,raise=-2pt}] (0,10) -- node[above=15pt] {Impurity coupling is ferromagnetic} (7,10);

 \draw [decorate,decoration={brace,amplitude=10pt,raise=-2pt}] (7,10) -- node[above=15pt] {Impurity coupling is antiferromagnetic} (14,10);
 
\end{tikzpicture}

    \caption{Phase diagram for Hamiltonian \eqref{xxximp} containing odd number of bulk sites. In the bound mode phase, the red curve is the energy of the bound mode, which exists in the ground state, and the blue line is the state in which the impurity is unscreened. In the ferromagnetic bound mode phase, the red line represents the ground state where the impurity is unscreened and the blue curve is the high-energy state containing the bound mode where the impurity spin is screened.}
    \label{fig:PD1}
\end{figure}
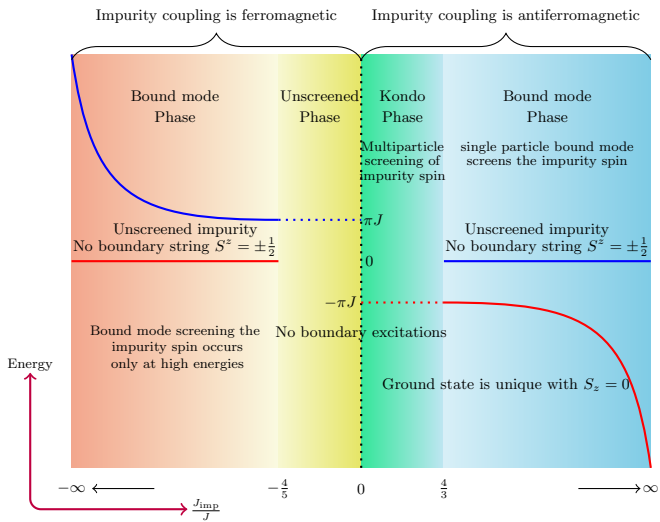

In the thermodynamic limit, the two impurities are independent (up to $1/L$ corrections). Thus, we find that each impurity exhibits the same phases as that of the one-impurity case, and hence the system exhibits sixteen different phases as shown in Fig \ref{PDeven}. However, the characteristics of the ground state and excitations are different, depending on the parity of the total number of sites and the ratio of the left and right boundary coupling to the bulk coupling as described in Sec~\ref{2imp}. We shall often parameterize the two boundary couplings as
\begin{equation}
    J_L=\frac{J}{1-b_L^2} \quad \text{ and }\quad  J_R=\frac{J}{1-b_R^2}
    \label{vardeff}
\end{equation}
as these are the natural variables for \textit{Bethe Ansatz equations} (see Appendix \ref{sec: hamder} and \ref{sec: DetailSol}). The parameters $b_L$ and $b_R$ can take either real values or purely imaginary values. As described above, one impurity exhibits four phases that are characterized by the ratio of couplings as shown in Fig.\ref{fig:PD1}. In terms of the parameter $b_i$, the Kondo phase exists when $b_i$ is purely imaginary or it takes real values $0<b_i<\frac{1}{2}$, the ABM phase exists when $\frac{1}{2}<b_i<1$, the FBM phase exists when $1<b_i<\frac{3}{2}$, and the unscreened phase exists when $b_i>\frac{3}{2}$.

The paper is organized as follows: Section \ref{BAresults} and \ref{dmrgresult} are dedicated to the one-impurity case. In section \ref{BAresults} we summarize the Bethe Ansatz results and discuss the impurity magnetization curves obtained using the density matrix renormalization group (DMRG) in section \ref{dmrgresult}. In section \ref{2imp} we consider the case of two impurities and discuss the structure of the ground states and the resulting phase diagram. Finally, we summarize our results in Section \ref{conc}.

\section{One Impurity: Bethe Ansatz results}\label{BAresults}
The exact ground state and the excitation spectrum depend on the relative values of the bulk and impurity coupling strengths and also depend on the parity of the number of sites in the system. Before discussing the results of our model, we briefly go over the ground states exhibited by the Heisenberg chain when open boundary conditions (OBC) are applied. This corresponds to the limit when the Kondo impurity coupling in our model is taken to zero. 

\smallskip

{\em The Heisenberg chain with OBC:}
For the spin chain with an even number of sites, the ground state is unique and it is a total singlet. It is represented by 

\begin{equation}
\ket{GS}\equiv \ket{0}\;\; \text{(even number of sites)}
\end{equation}

The ground state corresponding to the chain with an odd number of sites contains a spinon, which carries spin $\frac{1}{2}$. This spin can be oriented either in the positive or negative z-direction, which results in a two fold degenerate ground state, represented by

\begin{equation}
\ket{GS}\equiv \left|{\pm\frac{1}{2}}\right\rangle \;\; \text{(odd number of sites )}
\end{equation}

The energy of the spinon is given by
\begin{equation}
\label{espinon}
    E_{\theta}=\frac{2\pi J}{\cosh(\theta)},
\end{equation}
where $\theta$ is the rapidity. This indicates that the spinon in the ground state corresponding to the chain containing an odd number of sites has rapidity $\theta\rightarrow\infty$, which corresponds to the lowest energy of the spinon $E\rightarrow 0$ (in the thermodynamic limit). Note that the spinon has a maximum energy $T_0=2\pi J$, which corresponds to $\theta=0$.

\smallskip

We are now ready to summarize the Bethe Ansatz results, which are presented in detail in the Appendix \ref{sec: DetailSol}.

\subsection{Kondo phase}
\subsubsection{Ground state}

When the ratio of the impurity and bulk coupling strengths $(J_{\mathrm{imp}}/J)$ is less than $4/3$, the ground state for the chain with an odd number of bulk sites (even number of total sites) is a singlet, represented by

\begin{equation}
\ket{GS}\equiv \ket{0} \;\; \text{(odd number of bulk sites)}
\end{equation}

The ground state for the chain containing even number of bulk sites consists of a spinon with spin oriented in either the positive or negative z-direction and having rapidity $\theta\rightarrow\infty$, which corresponds to the lowest energy of the spinon \ref{espinon}. It is represented by

\begin{equation}
\ket{GS}\equiv \left|{\pm \frac{1}{2}}\right\rangle \;\; \text{(even number of bulk sites)}
\end{equation}

For both the odd and even number of site cases, the impurity is screened in the ground state. The screening occurs due to the genuine multiparticle Kondo effect which is discussed below.

The Kondo effect where the impurity is screened below a certain energy scale occurs when the impurity forms a many-body singlet with conduction electrons in a metal or with spins at various sites of the chain in our case. This effect is characterized by the appearance of a Lorentzian peak in the ratio of the impurity DOS ($\rho_{\mathrm{imp}}(E)$) and bulk DOS ($\rho_{\mathrm{bulk}}(E)$) : 
\begin{equation}
    R(E)= \frac{N}{2}\frac{{ \rho}_{\mathrm{imp}}(E)}{{ \rho}_{\mathrm{bulk}}(E)}
        \label{redef}
\end{equation}

The impurity DOS and the bulk DOS in the ground state are given by (see Appendix \ref{DOS})

\begin{equation}
    \rho_{\mathrm{imp}}(E)=\frac{\left(4 \pi ^2 J^2 \cos \left(\pi  b\right)\right)/\sqrt{4 \pi ^2 J^2-E^2}}{ \left(8 \pi ^2 J^2-E^2+E^2 \cos \left(2 \pi b\right)\right)},
\end{equation}
where we used the parameterization $\frac{J_{\mathrm{imp}}}{J}=\frac{1}{1-b^2}$ introduced earlier in Eq.\eqref{vardeff}
and \begin{equation}
    \rho_{\mathrm{bulk}}= \frac{N}{2 \pi \sqrt{4 \pi^2 J^2 - E^2}}.\end{equation}

%The ratio of the impurity to bulk density of state is given by
%\begin{equation}
    %R(E)=\frac{4 \pi ^2 J^2 \cos (\pi  \sqrt{1-\frac{J}{J_{\text{imp}}}})}{E^2 \cos (2 \pi  \sqrt{1-\frac{J}{J_{\text{imp}}}})-E^2+8 \pi ^2 J^2}
%\end{equation}

\begin{figure}
    \centering
    \includegraphics[width=\linewidth]{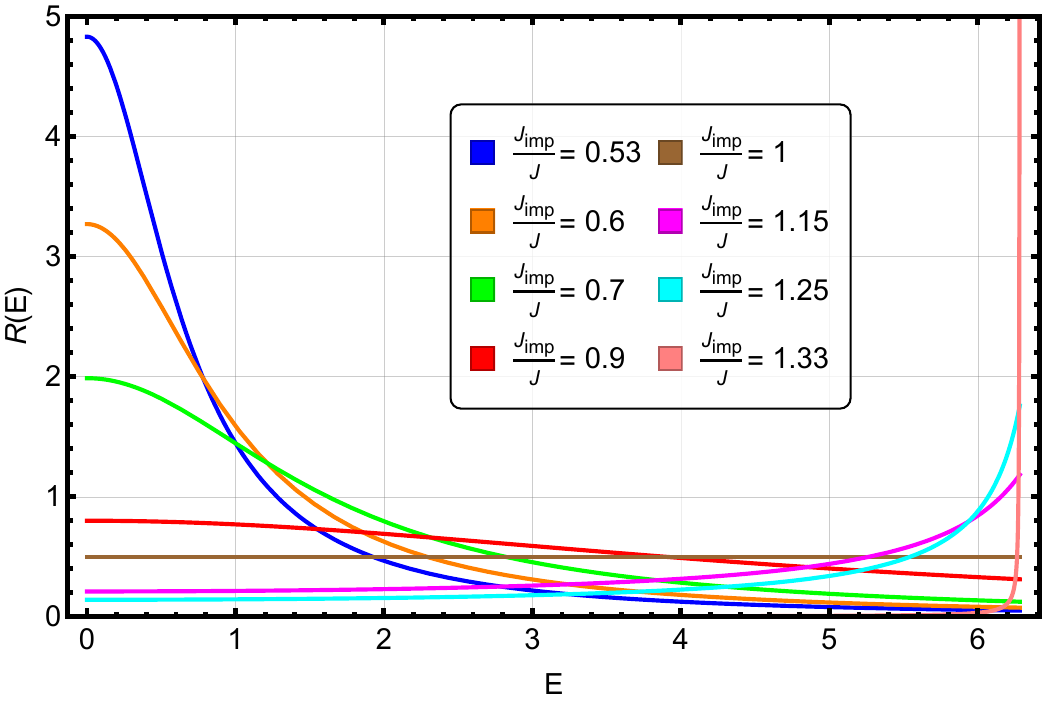}
    \caption{The plot offers a visualization of the spectral weight of the spinons in the Kondo cloud screening impurity. This spectral weight, denoted by the function $R(E)$ defined in Eq. \eqref{redef}, was computed using the \textit{Bethe Ansatz} method. Recall that the parameter $b$ in Eq.\eqref{redef} is defined as $\frac{J_{\mathrm{imp}}}{J}=\frac{1}{1-b^2}$. Notably, the plot reveals intriguing trends: deep within the Kondo regime, the majority of the spinons involved in impurity screening have energies close to the Fermi energy $(E=0)$. However, as the impurity coupling surpasses that of the bulk, a noteworthy shift occurs, with most of the screening spinons clustering around an energy level close to $2\pi J$, which represents the maximum energy of a single spinon. }
\label{fig:dos}
\end{figure}

Usually, when $R(E)$ takes a Lorentzian form, the Kondo temperature is defined as the width at half maximum of this Lorentzian peak. In our case, $R(E)$ takes a Lorentzian-like form when $J_{\mathrm{imp}}$ is much smaller than $J$, but undergoes a significant change as the impurity coupling strength increases. As $(J_{\mathrm{imp}}/J)$ approaches the value of $1$, it loses the Lorentzian form of the peak at $E=0$, and eventually takes the form of a constant function when $J_{\mathrm{imp}}=J$, where the impurity cannot be distinguished from the bulk. When the impurity coupling strength is further increased, a peak appears at $E=T_0$ and eventually takes the form of a delta function when $(J_{\mathrm{imp}}/J)=4/3$. See Fig [\ref{fig:dos}].

Due to the disappearance of the peak in the DOS, the above-described prescription to obtain the Kondo temperature does not work. Nevertheless, we find an equivalent way to obtain the Kondo temperature: It is the energy below which the number of states is exactly half of the total number of states associated with the impurity. This is expressed as

 \textit{i.e.}
\begin{equation}
    \int_0^{T_K} d E \rho_{i m p}(E)=\frac{1}{2} \int_0^{2\pi J} d E \rho_{i m p}(E).
\end{equation}
Evaluating this, we obtain
\begin{equation}
    T_K=\frac{2\pi J}{\sqrt{1+\cos^2(\pi b)}}.
    \label{tkdef}
\end{equation}

\begin{figure}
    \centering
    \includegraphics[width=\linewidth]{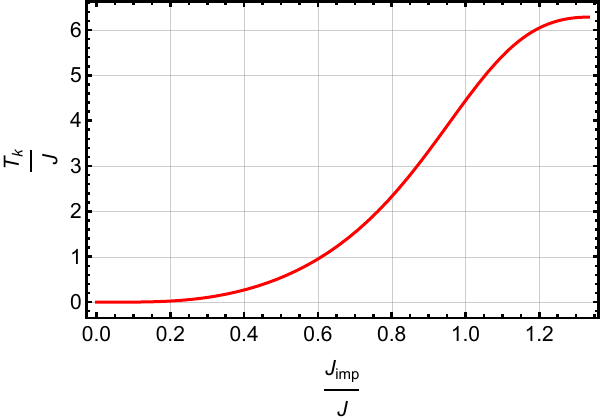}
    \caption{Within the Kondo regime, there is a characteristic Kondo temperature denoted as $T_K$. The expression for $T_K$ defined in Eq.\eqref{tkdef} is determined using the \textit{Bethe Ansatz} method and the parameter $b$ is defined in Eq.\eqref{vardeff}. In particular, as the impurity coupling value increases, the Kondo temperature grows. At the critical phase transition point, where $J$ equals $\frac{4}{3}$ of $J_{\mathrm{imp}}$, a significant connection arises: $T_K$ coincides with the energy of the bound mode, represented as $E_b$ given by Eq.\eqref{EbMeng}.
    }
    \label{fig:enter-label}
\end{figure}

The Kondo temperature as a function of $(J_{\mathrm{imp}}/J)$ is shown in Fig [\ref{fig:enter-label}].

Deep in the Kondo regime $J_{\mathrm{imp}}\ll J$, the Kondo temperature scales as
\begin{equation}
    T_K\sim e^{-\frac{1}{2} \pi  \sqrt{\frac{J}{\mathrm{Jimp}}}}
\end{equation}
The square root behavior of $T_K$ in the exponent is unlike the Fermi liquid Kondo problem where $T_k\sim e^{-\frac{1}{J_{imp}}}$. This result has been reported in previous studies \cite{wang1997exact,frahm1997open,laflorencie2008kondo}. However, our definition of $T_K$ differs from all previously obtained results for the higher value of the impurity coupling. The $T_K$ we obtained is continuous in all parametric regimes of the Kondo phase and at the phase transition line it is exactly equal to the energy of the bound mode which sets the scale of the impurity physics in the bound mode phase.
\subsubsection{Excitation spectrum}
The excited states in the Kondo phase can be constructed above the ground state by adding even number of spinons, bulk strings, quartets, etc. \cite{destri1982analysis} which form a tower. Note that since the ground state for even number of sites contains a spinon, whereas the ground state for the chain with an odd number of sites does not, in any excited state the number of spinons is always odd in the chain with even number of sites, whereas it is always even for the chain with an odd number of sites. 

\subsection{ABM phase}
As was discussed in the previous subsection, the ratio of DOS $R(E)$ takes the form of a delta function peak centered at $E=T_0$ when the ratio of the impurity and the bulk coupling strengths approaches $4/3$. This suggests that the many body screening effect in the Kondo phase essentially turns into a single particle screening when $(J_{\mathrm{imp}}/J)=4/3$. We find that as this ratio is increased beyond the value of $4/3$, the ground state contains an exponentially localized bound state that is described by the boundary string:

\begin{equation}
    \mu_b=\pm i\left(\frac{1}{2}-b\right),
    \label{bssoln}
\end{equation}

which screens the impurity. This results in the ground state for the chain with an odd number of bulk sites to have spin $S^z=0$ which is represented by

\begin{equation}
\ket{GS}\equiv \ket{0}.   \;\; \text{(odd number of bulk sites)}
\end{equation}

Analogous to the Kondo phase, the ground state for the chain with even number of bulk site consists of a spinon with spin oriented in the positive or negative $z$-direction with rapidity $\theta\rightarrow\infty$, which is represented by

\begin{equation}
\ket{GS}\equiv \left|{\pm\frac{1}{2}}\right\rangle. \;\; \text{(even number of bulk sites)}
\end{equation}

\subsubsection{Excitation spectrum}
As described above, unlike in the Kondo phase, the system exhibits a bound state described by the boundary string. This boundary string can be removed and hence one can un-screen the impurity, which now costs energy that depends on the ratio of the impurity and bulk coupling strengths:

\begin{equation}\label{bmenergy}
    E_b=\frac{2\pi J}{\sin\left(\pi b\right)}.
\end{equation}

Notice that this is always above the maximum energy of the single spinon $T_0$, and goes to infinity as $(J_{\mathrm{imp}}/J)\rightarrow \infty$. 

For the chain with an even number of bulk sites, the resulting state in which the impurity is unscreened has spin $S^z=\pm\frac{1}{2}$ which corresponds to the spin of the unscreened impurity. It is represented by

\begin{equation}
  \ket{US}\equiv \ket{\pm\frac{1}{2}}. \;\; \text{(even number of sites bulk chain)}  
\end{equation}

For the chain with an odd number of bulk sites, the state in which the impurity is unscreened contains a spinon, whose spin can be oriented either along or opposite to that of the impurity, and as a result, this state is four-fold degenerate and is represented by

\begin{equation}
\ket{US}\equiv \ket{\pm 1}, \ket{0}, \ket{0}'. \;\; \text{(odd number of bulk sites)}   
\end{equation}

Here $\ket{0},\ket{0}'$ are symmetric and anti-symmetric under the exchange of spin of the impurity and the spinon respectively.

Starting from either the ground state in which the impurity is screened or from the state in which the impurity is unscreened, one can build up excitations in the bulk by adding even number of spinons, bulk strings, quartets and wide boundary strings, and one obtains two towers of excited states. 

Note that just like in the Kondo phase, in any excited state corresponding to the tower in which the impurity is screened, the number of spinons corresponding to the chain with an even or odd number of sites is always odd or even respectively. 

Since the lowest energy state corresponding to the tower in which the impurity is unscreened consists of a spinon for the chain with an odd number of sites, whereas it is absent in the chain with an even number of sites , all the excited states corresponding to this tower have odd or even number of spinons for a spin chain with an odd or even number of sites respectively.

\subsection{FBM and US phases}
When the impurity coupling is Ferromagnetic, the impurity is always unscreened in the ground state. For the chain consisting of an even number of bulk sites, the ground state is two fold degenerate with spin $Sz=\pm\frac{1}{2}$ corresponding to the spin of the impurity. It is represented by

\begin{equation}
\ket{GS}=\left|{\pm\frac{1}{2}}\right\rangle.\;\; \text{(even number of bulk sites)}
\end{equation}

When the number of bulk sites is odd, the ground state contains a spinon, whose spin can be oriented either along or opposite to that of the impurity, and as a result, this state is four fold degenerate. It is represented by

\begin{equation}
\ket{GS}\equiv \ket{\pm 1}, \ket{0}, \ket{0}' \;\; \text{(odd number of bulk sites)}   
\end{equation}

where $\ket{0},\ket{0}'$ are symmetric and anti-symmetric under the exchange of spin of the impurity and the spinon respectively.

\subsubsection{Excitation spectrum: FBM phase}

When the ratio of the impurity and bulk coupling strengths is greater than $4/5$, there exists an exponentially localized bound mode described by the boundary string \ref{bssoln}, which can be added to the ground state and thereby screen the impurity. The energy of the bound mode is given by \ref{bmenergy}. 

For the chain with an odd number of bulk sites, this bound mode can be added to the ground state by removing the existing spinon, resulting in a state with spin $S^z=0$ represented by

\begin{equation}
    \ket{S}\equiv \ket{0}.\;\; \text{(odd number of bulk sites)}
\end{equation}
 
 For the chain with even number of bulk sites, the bound mode can be added by adding a spinon whose spin is oriented in either the positive or negative $z$-direction, resulting in a two fold degenerate state with spin $S^z=\pm \frac{1}{2}$ represented by

 \begin{equation}
     \ket{S}\equiv \left|{\pm\frac{1}{2}}\right\rangle.\;\; \text{(even number of bulk sites)}
 \end{equation}

Starting with either the ground state in which the impurity is unscreened or with the state in which the impurity is screened, one can build up excitations in the bulk by adding even number of spinons, bulk strings, quartets and wide boundary strings, and one obtains two towers of excited states.

Note that unlike the antiferromagnetic bound mode phase, in any excited state corresponding to the tower in which the impurity is screened, the number of spinons corresponding to the chain with an even or odd number of sites is always even or odd respectively. Whereas, all the excited states corresponding to the tower in which the impurity is unscreened have odd or even number of spinons for a spin chain with odd or even number of sites respectively.

\subsubsection{Excitation spectrum: US phase}
Unlike in the ferromagnetic bound mode phase, the impurity cannot be screened, and hence the Hilbert space contains a single tower with excitations built on top of the ground state.
All the excited states in this tower have an odd or even number of spinons for a spin chain with an odd or even number of sites respectively.

\section{One impurity: Local impurity magnetization}\label{dmrgresult}

In this section we use DMRG technique to compute the impurity magnetization in the presence of a global magnetic field. 
%$-h\sum_{j}\sigma_{j}$, 
%where the summation index extends over bulk and impurity sites. 
%
%
\begin{figure*}[ht!]
    \begin{minipage}{0.33\textwidth}
        \includegraphics[height=\linewidth]{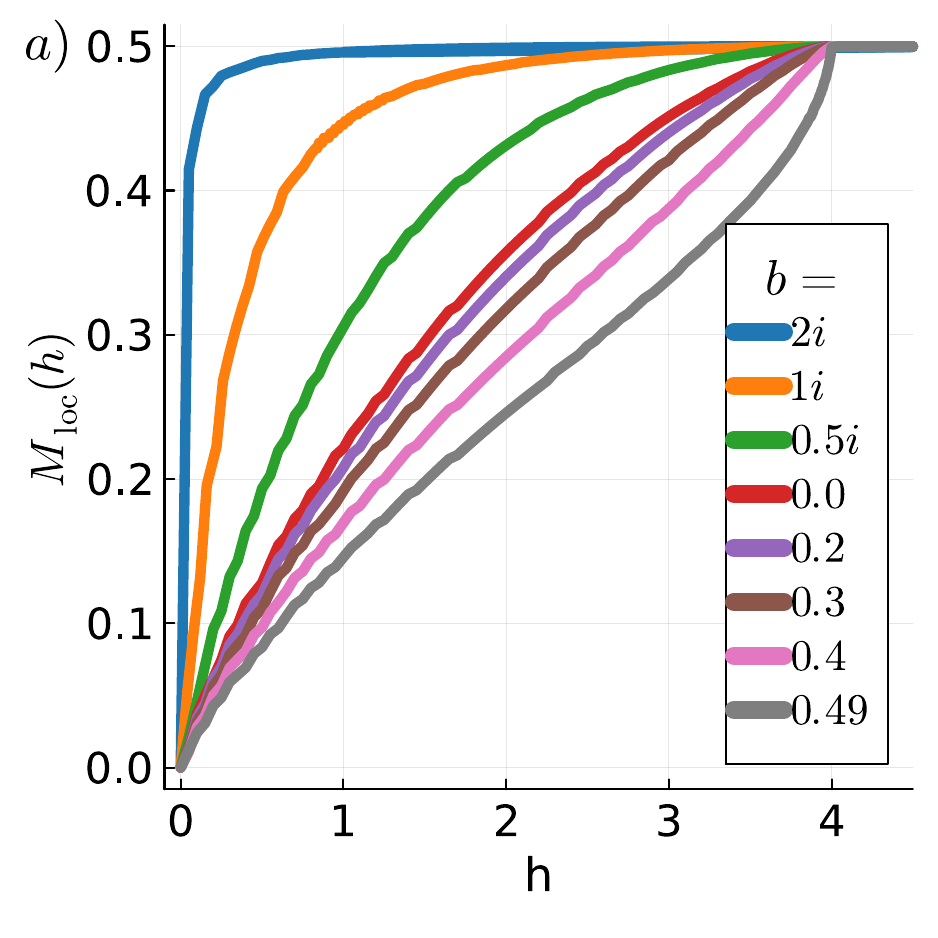}
    \end{minipage}%
    \begin{minipage}{0.33\textwidth}
        \includegraphics[height=\linewidth]{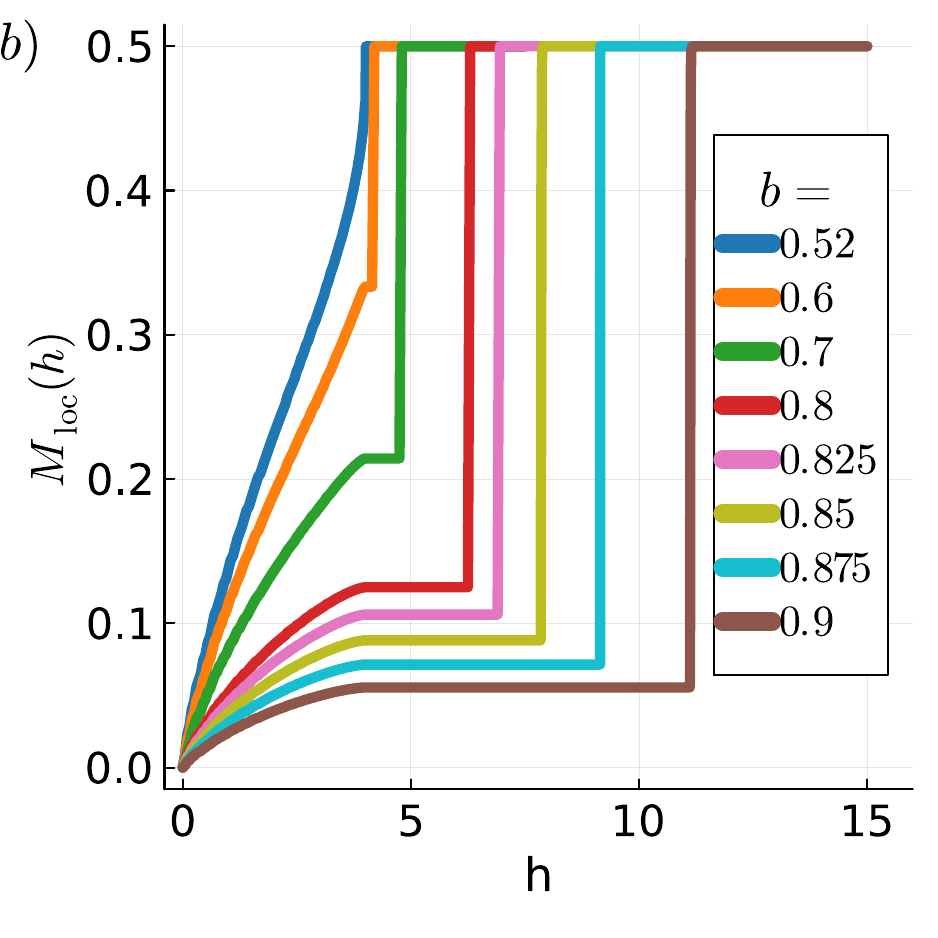}
    \end{minipage}%
    \begin{minipage}{0.33\textwidth}
        \includegraphics[height=\linewidth]{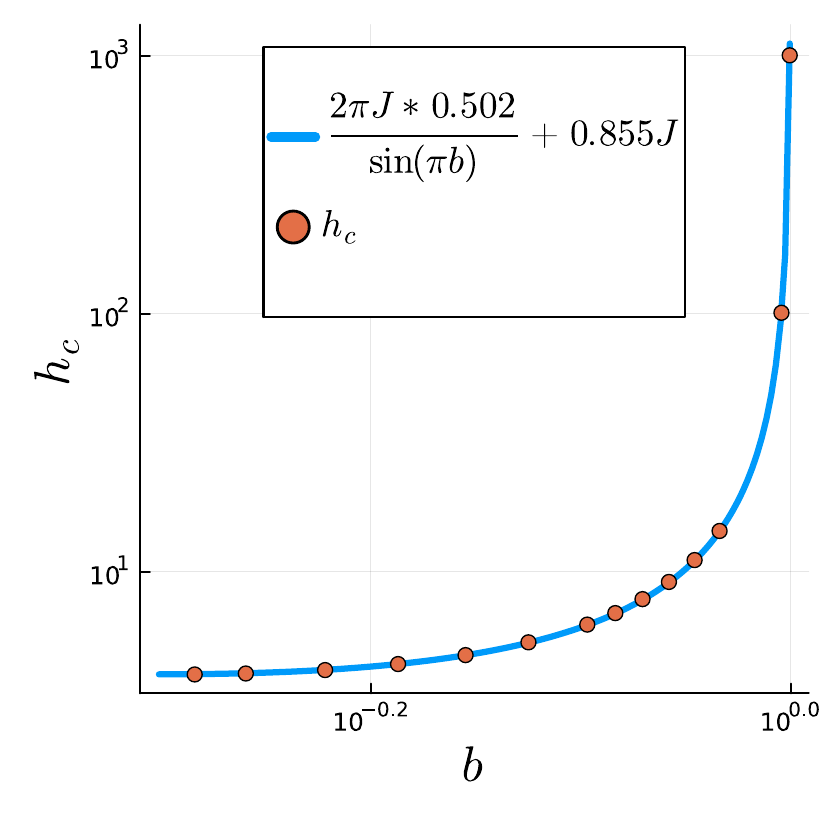}
    \end{minipage}
\caption{Local impurity magnetization defined in Eq.~\eqref{eqn:Mloc} computed using DMRG. All calculations are performed for $N = 499$ bulk sites and 1 impurity, a truncation error cutoff set at $10^{-10}$ and 50 sweeps to ensure the reliability and convergence of our calculations. a)Local impurity magnetization for various values of coupling in the Kondo phase (purely imaginary $b$ or $0 < b < \frac{1}{2}$. The impurity magnetization transitions smoothly from 0 to $\frac12$  as the external magnetic field increases which is consistent with the Kondo-like behavior. When $h=4J$, all the spins in the XXX chain are fully polarized \cite{granet2019analytical} and so does the impurity in the Kondo phase. b) Local impurity magnetization for various values of coupling in the bound mode phase ($\frac{1}{2} < b < 1$). The magnetization smoothly increases to some values for $h<4J$. At $h=4J$ all other spins in the spin chain are fully polarized, but the impurity magnetization is constant up until it saturates jumps when the magnetic field is equal to the renormalized energy of the bound mode in the presence of the magnetic field.  c) In the bound mode phase, the critical value of field at which magnetization jumps gives the energy of bound mode in the presence of magnetic field. We fit the critical value of magnetic field to a function 
 $h_c = h_0 +x|E_b|$ and extract $h_0 = 0.855$J and $x\approx 0.5$. }
\label{fig:impmaglargeb}
\end{figure*}
In order to show that the local physical quantities behave differently in the Kondo and bound mode physics via DMRG, we resort to Hamiltonian \eqref{xxximp} and rewrite it by adding the global magnetic field $h$ as
\begin{equation}
    H=\sum_{j=1}^{N-1}  \left(J \vec\sigma_j \cdot \vec\sigma_{j+1}
    %-h\sum_j \sigma^z_j
    -h \sigma^z_j \right)
    +\frac{J}{1-b^2}\vec\sigma_1\cdot\vec \sigma_\mathrm{imp}- h\sigma^z_{\mathrm{imp}}
    %\nonumber
\end{equation}
where we use the parameterization introduced in Eq.\eqref{vardeff} $J_{\mathrm{imp}}=\frac{J}{1-b^2}$ such that the bound mode phase is when $\frac{1}{2}<b<1$ and the Kondo phase is when $0<b<\frac{1}{2}$ and also when $b$ is purely imaginary as mentioned earlier.

Taking a spin chain of total $N=500$ sites (499 bulk sites and one impurity at the left end), for various values of $h$, we compute the local magnetization at the impurity site

\begin{equation}
%\mathcal{M}_i
M_{\mathrm{loc}}(h)=\langle \sigma_{\mathrm{imp}}^z\rangle
\label{eqn:Mloc}
\end{equation}
and compare the results in Kondo and bound mode phases. The DMRG calculation of this local impurity magnetization is performed using the ITensor library \cite{fishman2022itensor}. All DMRG calculations are performed with a truncation error cutoff of $10^{-10}$. Depending on the value of magnetic field and the boundary coupling, the calculation converges after different number of sweeps. Thus, to ensure the convergence for all ranges of parameter, we used 50 sweeps for all calculation. 
In the Kondo phase, we see a smooth crossover from $M_{\mathrm{loc}}(h=0)=0$ as the applied field is zero to $M_{\mathrm{loc}}(h)=\frac{1}{2}$ at $h_c=4J$ reminiscent of the Fermi-liquid Kondo problem \cite{andrei1983solution}. However, in the bound mode phase, the impurity magnetization ($M_{\mathrm{loc}}(h)$) grows smoothly only up until $h=4J$ to a finite values smaller than $\frac{1}{2}$ and then saturates. Once the magnetic field has energy equal to that of the bound mode (in the presence of the magnetic field), the impurity is unscreened. Thus, at some finite value of $h$, the impurity magnetization $M_{\mathrm{loc}}(h)$  abruptly jumps from a finite value to $\frac{1}{2}$, thereby proving the existence of the bound mode. 

We compute the impurity magnetization for a range of parameter $b$ in the Kondo phase. Recall that the Kondo phase is where $0<(J_{\mathrm{imp}}/J)<4/3$ or equivalently where the paramter $b$ takes either purely imaginary values or real values between $0$ and $\frac{1}{2}$.
%
% \begin{figure}[H]
%     \centering
%     \includegraphics[width=\linewidth]{images/kondo.pdf}
%     \caption{Local impurity magnetization for various values of coupling in the Kondo phase (purely imaginary b or $0<b<\frac{1}{2}$) for $N=500$ sites is computed using DMRG. We implemented a truncation error cutoff set at $10^{-10}$ and performed 50 sweeps to ensure the reliability and convergence of our calculations. The impurity magnetization transitions smoothly from 0 to $\frac{1}{2}$ as the external magnetic field increases which is consistent with the Kondo-like bahavior.}
%     \label{fig:impmagkondo}
% \end{figure}
%
%
Notice that for imaginary values of $b$ ($J_{\mathrm{imp}}<J$, the magnetization curve is a concave upward increasing curve just like in the Fermi-liquid Kondo model. Here, the impurity magnetization asymptotically reaches the free spin limit $\frac{1}{2}$. 
However, the shape changes to concave upward increasing as $b$ becomes real (i.e.$\frac{4}{3}J>J_{\mathrm{imp}}>J$. The concave upward nature of the graph is most prominent for $b=0.49$ in Fig.\ref{fig:impmaglargeb}(a) which is near the phase transition line. In this regime $0<b<\frac{1}{2}$, the magnetization is not exactly Fermi-liquid Kondo-like as the magnetization curve has an upturn to reach the free impurity value of$\frac{1}{2}$ which is most likely the effect of shift in Kondo peak from the $E=0$ to $E=2\pi J$ as discussed above. 

Now, we compute the impurity magnetization in the bound-mode phase. Recall that the bound mode energy in terms of the parameter $b$ is
\begin{equation}
    E_b =-\frac{2\pi J}{\sin(\pi b)}
    \label{EbMeng}
\end{equation}
where $\frac{1}{2}<b<1$. As $b\to 1$, both the impurity coupling and the energy of the bound mode tend to infinity. We compute the impurity magnetization for several values of $b$. We observe that for small values of $h<4j$, impurity magnetization grows just like in the Kondo phase, but then it saturates for intermediate $h$, and finally jumps when it has enough energy to break apart the bound mode. 

% \begin{figure}[H]
%  \includegraphics[width=\linewidth]{images/bmmag.pdf}
%  \caption{Local impurity magnetization in bound mode phase $\left(\frac{1}{2}<b<1\right)$ for $N=500$ sites chain is computed using DMRG. The DMRG calculation is performed with truncation cutoff of $10^{-10}$ and 50 sweeps. The magnetization jumps due to the existence of the bound mode. The value of $h$ at which the jump occurs is related to the energy of the bound mode.  }
%  \label{bmimpmag}
% \end{figure}
% \begin{figure}[H]
%     \centering
%     \includegraphics[width=\linewidth]{images/pltfit.pdf}
%     \caption{Critical value of field at which magnetization jumps which are read off the plot FIG.~\ref{bmimpmag}. We fit the critical value of magnetic field to a function $h_c=h_0+x|E_b|$ and extract $h_0=1.71J$ and $x\approx 1$. 
%     }
%     \label{fig:impmaglargeb}
% \end{figure}

We observe that as $b$ increases the initial growth of impurity magnetization is supressed, the  magnetization pleatues and finally when the magnetic field has enough energy to overcome the bound mode energy, the magnetization jumps to $\frac{1}{2}$ which is consistent with the fact that there exists massive bound mode in the ground state and hence this regime is not observed in the low energy boundary CFT description.

Notice that the energy of bound mode $E_b=\frac{2\pi J }{\sin(\pi b)}$ was computed for $h=0$. When the external magnetic field $h$ is non-zero, the energy of the bound mode will change. The critical magnetic field, where the impurity magnetization jumps, is equal to the energy of the bound mode in the presence of the magnetic field, serving as a measure of the boundary gap in the model.
We propose that the critical field is related to the energy of the bound mode for zero magnetic field $E_b$ as
\begin{equation}
    h_c=h_0+x|E_b|
\end{equation}

We compute the critical value of the field at which the impurity magnetization jumps for various values of the parameter $b$ and by fitting the data with $h_c=h_0 + x |E_b|$, we find that it satisfies

\begin{equation}
    h_c=0.855 J+ 0.502 |E_b|
\end{equation}
as shown in FIG.~\ref{fig:impmaglargeb}.

\definecolor{dy}{rgb}{0.9,0.9,0.5}
\definecolor{dr}{rgb}{0.95,0.55,0.5}
\definecolor{db}{rgb}{0.5,0.8,0.9}
\definecolor{dg}{rgb}{0.3,0.9,0.75}
\definecolor{dm}{rgb}{0.3,0.6,0.85}
\definecolor{dn}{rgb}{1,0.7,0.6}

\begin{figure}
\begin{center}
\begin{tikzpicture}[scale=0.6, every node/.style={scale=0.7}]
\draw [<->, rounded corners, thick, purple] (-13,-10) -- (-13,-13) -- (-10,-13);
\node at (-13,-9.75) {$b_R$};
\node at (-9.75,-13) {$b_L$};
\draw [thick] (-12,0)--(-12,-12) -- (0,-12);
\draw [thick] (-10,-12)--(-10,0);
\draw [thick] (-12,0)--(0,0)--(0,-12);
\draw [thick] (-2,-12)--(-2,0);
\draw [thick] (-12,-2)--(0,-2);
\draw [thick] (-12,-10)--(0,-10);
\draw [fill=dg] (-10,-10)--(-10,-2)--(-2,-2)--(-2,-10);

\draw [fill=dm] (-12,-12)--(-12,-10)--(-10,-10)--(-10,-12);
\draw [fill=dr] (-2,-2)--(-2,0)--(0,0)--(0,-2)--(-2,-2);
\draw [fill=dy] (-10,-2)--(-10,0)--(-12,0)--(-12,-2)--(-10,-2);
\draw [fill=dy] (0,-12)--(0,-10)--(-2,-10)--(-2,-12)--(0,-12);
\draw [fill=db] (-12,-10)--(-12,-2)--(-10,-2)--(-10,-10)--(-12,-10);
\draw [fill=dn] (-10,-2)--(-10,0)--(-2,0)--(-2,-2)--(-10,-2);
\draw [fill=dn] (-2,-2)--(0,-2)--(0,-10)--(-2,-10)--(-2,-2);
\draw [fill=db] (-10,-12)--(-10,-10)--(-2,-10)--(-2,-12)--(-10,-12);

\draw [thin] (-6,-12)--(-6,0);
\draw [thin] (-10,-12)--(-10,0);
\draw [thin] (-12,-6)--(0,-6);

\node at (-4,-1) {\scriptsize{$(FBM-US)$}};

\node at (-8,-1) {\scriptsize{$(BM-US)$}};

\node at (-4,-11) {\scriptsize{$(FBM-K)$}};

\node at (-8,-11) {\scriptsize{$(ABM-K)$}};

\node at (-1,-11) {\scriptsize{$(US-K)$}};

\node at (-11,-11) {\scriptsize{$(K-K)$}};

\node at (-1,-1) {\scriptsize{$(US-US)$}};

\node at (-11,-1) {\scriptsize{$(K-US)$}};

\node at (-11,-8) {\scriptsize{$(K-ABM)$}};

\node at (-11,-4) {\scriptsize{$(K-FBM)$}};

\node at (-1,-8) {\scriptsize{$(US-ABM)$}};

\node at (-1,-4) {\scriptsize{$(US-FBM)$}};

\node at (-12.4,-6) {1};
\node at (-12.4,-10) {$\frac12$};
\node at (-12.4,-2) {$\frac{3}{2}$};
\node at (-2,-12.4) {$\frac{3}{2}$};
\node at (-6,-12.4) {1};
\node at (-10,-12.4) {$\frac12$};

\node at (-4,-8) {\scriptsize{$(FBM-ABM)$}};

\node at (-4,-4) {\scriptsize{$(FBM-FBM)$}};

\node at (-8,-4) {\scriptsize{$(ABM-FBM)$}};

\node at (-8,-8) {\scriptsize{$(ABM-ABM)$}};

\end{tikzpicture}

\end{center}
\caption{Phase diagram for even chain when there are two impurities. There are 16 phases in total, which is due to there being 4 independent phases at each end, as shown in Fig.\ref{fig:PD1}.}
\label{PDeven}
\end{figure}
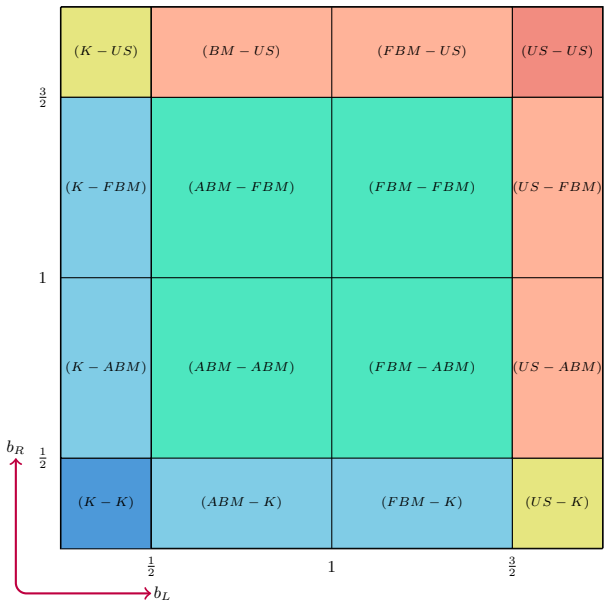 

\section{Two Impurities: Bethe Ansatz Results}\label{2imp}
In this section, we consider the case of two spin $\frac{1}{2}$ impurities interacting with the two edges of the chain with coupling strengths $J^i_{\mathrm{imp}}, \; i=\{L,R\}$. As mentioned previously, this system for arbitrary spin $S$ of the impurities was shown to be integrable for arbitrary values of the impurity coupling strengths \cite{wang1997exact}, and was solved for the space parity symmetric case $J^L_{\mathrm{imp}}=J^R_{\mathrm{imp}}$\cite{wang1997exact}. Here, we provide the solution for arbitrary values of the impurity couplings. We find that each impurity exhibits the same four phases as discussed above, and hence the two impurity system exhibits a total of sixteen phases as shown in the Fig \ref{PDeven}. Here, we used the parameters $b$ and $d$ instead of the ratio of the bulk and boundary couplings, where as defined in Eq.\eqref{vardeff} $J_L=\frac{J}{1-b_L^2}$ and $J_R=\frac{J}{1-b_R^2}$.

%One can define the bound state parities associated with each edge 

%\begin{equation}\mathcal{P}_{L,R}=(-1)^{N_{L,R}}\end{equation}

%where $N_{L,R}={0,1}$ corresponds to number of bound states at the left and the right edges respectively. 

The explicit construction of the ground states and the elementary excited states in each of the sixteen phases is shown in the appendix \ref{appendix}. Here, we briefly summarize the Bethe Ansatz results in various phases.

\subsection{Kondo-Kondo phase}
This phase occurs when the ratio of the impurity and the bulk coupling strengths for both impurities is less than $\frac{4}{3}$. Both impurities are screened in the ground state through the Kondo effect, which gives rise to two Kondo temperatures $T_{Ki}, i=\{L,R\}$, where

\begin{equation}
    T_{Ki}=\frac{2\pi J}{\sqrt{1+\cos^2(\pi b_i)}}, \;\; i=\{L,R\}.
    \label{tk2imp}
\end{equation}

The ground state for the chain with an even number of sites is unique, and it is a total singlet:

\begin{equation}\ket{GS}\equiv \ket{0}. \;\; \text{(even number of sites)}\end{equation}

The ground state for chain with odd number of sites contains a spinon with rapidity $\theta\rightarrow\infty$, whose spin is oriented either in the positive or negative $z$-direction, resulting in a two-fold degenerate ground state:

\begin{equation}\ket{GS}\equiv\left|{\pm\frac{1}{2}}\right\rangle.\;\; \text{(odd number of sites )}\end{equation}

\subsection{Kondo-ABM and ABM-Kondo phases}
Kondo-ABM phase occurs when the ratio of the impurity and bulk coupling strengths corresponding to the left and right impurities is $0<(J^{L}_{\mathrm{imp}}/J)<\frac{4}{3}$ and $(J^{R}_{\mathrm{imp}}/J)>\frac{4}{3}$ respectively.

In the ground state corresponding to the Kondo-ABM phase, the impurity coupled to the left edge is screened by the Kondo effect giving rise to the Kondo temperature $T_{KL}$, whereas the impurity coupled to the right edge is screened by an exponentially localized bound state.

The ground state for the chain containing even number of sites is unique and has total spin $S^z=0$, which is represented by

\begin{equation}\ket{GS}\equiv \ket{0}.\;\; \text{(even number of sites)}\end{equation}

Similar to the Kondo-Kondo phase, the ground state for the chain with an odd number of sites in the Kondo-ABM phase contains a spinon with rapidity $\theta\rightarrow\infty$, whose spin is oriented either in the positive or negative $z$-direction, resulting in a two fold degenerate ground state:

\begin{equation}\ket{GS}\equiv\left|{\pm\frac{1}{2}}\right\rangle.\;\; \text{(odd number of sites )}\end{equation}

The ABM-Kondo phase is related to Kondo-ABM phase by space parity transformation. The above described results follow through by applying the transformation $L\leftrightarrow R$.

The bound state screening the left or right impurity in the phases ABM-Kondo or Kondo-ABM phases respectively can be removed, and thereby unscreen the respective impurities. This requires one to add or remove the spinon with rapidity $\theta\rightarrow\infty$ for the spin chain with an even or odd sites, respectively. This costs energy  

\begin{equation}\label{bmenergy2imp}
    E^i_b=\frac{2\pi}{\sin\left(\pi b_i\right)},
\end{equation}

where $i=\{L,R\}$ for ABM-Kondo and Kondo-ABM phases respectively.

\subsection{Kondo-FBM and FBM-Kondo phases}
Kondo-FBM phase occurs when the ratio of the impurity and bulk coupling strengths corresponding to the left and right impurities is $0<(J^L_{\mathrm{imp}}/J)<\frac{4}{3}$ and $(J^R_{\mathrm{imp}}/J)<-\frac{4}{5}$ respectively.

In the ground state corresponding to the Kondo-FBM phase, the impurity coupled to the left edge is screened by the Kondo effect, giving rise to the Kondo temperature $T_{KL}$, whereas the impurity coupled to the right edge is unscreened.

The ground state for the chain with an even number of sites contains a spinon with rapidity $\theta\rightarrow\infty$, and as a result it is four fold degenerate:

\begin{equation}\ket{GS}\equiv\ket{\pm 1}, \ket{0},\ket{0}',\;\; \text{(even number of sites)}\end{equation}

where $\ket{0},\ket{0}'$ are symmetric and antisymmetric under the exchange of the spins of the impurity and the spinon.

The ground state for the chain with an odd number of sites in the Kondo-FBM phase is two-fold degenerate with spin $S^z=\pm\frac{1}{2}$ corresponding to the spin of the unscreened impurity. It is represented by

\begin{equation}\ket{GS}\equiv\left|{\pm\frac{1}{2}}\right\rangle.\;\; \text{(odd number of sites )}\end{equation}

The FBM-Kondo phase is related to the Kondo-FBM phase by space parity transformation. The above described results follow through by applying the transformation $L\leftrightarrow R$.

A bound state can be added and thereby screen the left and right impurities in the FBM-Kondo and Kondo-FBM phases respectively. This requires adding or removing the spinon with rapidity $\theta\rightarrow\infty$ for the chain with an odd and even number of sites, respectively. This costs energy $E^{L}_b,E^{R}_b$ for FBM-Kondo and Kondo-FBM phases respectively.

\subsection{Kondo-US and US-Kondo phases}
Kondo-US phase occurs when the ratio of impurity and bulk coupling strength corresponding to left and right impurities is $0<(J^L_{\mathrm{imp}}/J)<\frac{4}{3}$ and $0>(J^R_{\mathrm{imp}}/J)>-\frac{4}{5}$ respectively. The ground state in this phase is the same as that in the Kondo-FBM phase. There exist no bound states that can be added to screen the impurity. The US-Kondo phase is related to the Kondo-US phase by the space parity transformation.

\subsection{ABM-ABM phase}
This phase occurs when the ratio of the impurity and bulk coupling strengths corresponding to both impurities is $0<(J^i_{\mathrm{imp}}/J)<\frac{4}{3}, i=\{L,R\}$. 

In the ground state, both impurities are screened because of the exponentially localized bound modes. The ground state for the chain with an even number of sites has total spin $S^z=0$ and is represented by

\begin{equation}
    \ket{GS}\equiv \ket{0}. \;\; \text{(even number of sites)}
\end{equation}

The ground state for the chain with an odd number of sites has a spinon with rapidity $\theta\rightarrow\infty$ and whose spin is oriented either in the positive or negative $z$-direction, resulting in a two-fold degenerate ground state:

\begin{equation}\ket{GS}\equiv\left|{\pm\frac{1}{2}}\right\rangle.\;\; \text{(odd number of sites )}\end{equation}

The bound state screening either of the impurities can be removed by adding or removing the spinon with rapidity $\theta\rightarrow\infty$ for even and odd numbers of site chains, respectively. This costs energy $E^{L}_b,E^{R}_b$ to unscreen the left and right impurities respectively.

One can also remove both the bound states thereby unscreen both the impurities without having to remove or add any spinons. This costs energy $E^L_b+E^R_b$.

\subsection{FBM-FBM phase}
This phase occurs when the ratio of the impurity and bulk coupling strengths corresponding to both the impurities is $0>(J^i_{\mathrm{imp}}/J)>-\frac{4}{5}, i=\{L,R\}$. 

In the ground state, both impurities are unscreened. The ground state for even number of site chain is four-fold degenerate:

\begin{equation}
    \ket{GS}\equiv \ket{\pm 1}, \ket{0},\ket{0}', \;\; \text{(even number of sites)}
\end{equation}

where $\ket{0},\ket{0}'$ are symmetric and antisymmetric under the exchange of spins of both impurities. The ground state for the the chain with odd number of sites has a spinon with rapidity $\theta\rightarrow\infty$, and whose spin is oriented either in the positive or negative $z$-direction, resulting in an eight fold degenerate ground state corresponding to the spin multiplets of the highest weight states:

\begin{equation}\ket{GS}\equiv\left|{\frac{3}{2}}\right\rangle,\left|{\frac{1}{2}}\right\rangle,\left|{\frac{1}{2}}\right\rangle'.\;\; \text{(odd number of sites )}\end{equation}

Here, $\ket{\frac{1}{2}},\ket{\frac{1}{2}}'$ are symmetric and antisymmetric under the exchange of spins of the unscreened impurities.

Either of the impurities can be screened by adding or removing the spinon with rapidity $\theta\rightarrow\infty$ for the chain that contains an even and odd number of sites, respectively. This costs energy $E^{L}_b,E^{R}_b$ to screen the left and right impurities respectively. One can also add both the bound states and thereby screen both the impurities without having to remove or add any spinons. This costs energy $E^L_b+E^R_b$.

\subsection{ABM-FBM and FBM-ABM phases}

ABM-FBM phase occurs when the ratio of the impurity and bulk coupling strengths corresponding to the impurities at the left and right edges is $0<(J^L_{\mathrm{imp}}/J)<\frac{4}{3}$ and  $0>(J^R_{\mathrm{imp}}/J)>-\frac{4}{5}$ respectively. 

In the ground state, the left impurity is screened due to the presence of a bound state, whereas the impurity at the right edge is unscreened. The ground state for the chain with an even number of sites  consists of a spinon with rapidity $\theta\rightarrow\infty$, and whose spin can be oriented either in the positive or negative $z$-direction, resulting in a four fold degenerate ground state:

\begin{equation}
    \ket{GS}\equiv \ket{\pm 1}, \ket{0},\ket{0}'. \;\; \text{(even number of sites)}
\end{equation}

Here $\ket{0},\ket{0}'$ are symmetric and antisymmetric under the exchange of the spins of both impurities. The ground state for odd number of sites is two fold degenerate with spin $S^z=\pm\frac{1}{2}$, corresponding to the spin of the unscreened impurity. It is represented by

\begin{equation}\ket{GS}\equiv\left|{\pm\frac{1}{2}}\right\rangle.\;\; \text{(odd number of sites )}\end{equation}

The impurity at the left edge can be unscreened by removing the bound state. This requires one to remove or add the spinon with rapidity $\theta\rightarrow\infty$ in the ground state for the chain with an even or odd number of sites, respectively, which costs energy $E^L_b$. Similarly,  The impurity at the right edge can be screened by adding the bound state, which requires adding or removing a spinon with rapidity $\theta\rightarrow\infty$ in the ground state for the chain containing an even or odd number of sites respectively, which costs energy $E^R_b$. 

One can simultaneously unscreen and screen the impurities at the left and right edges by removing and adding the corresponding bound states, without having to add or remove spinons in the ground state. This costs energy $E^L_b+E^R_b$.

The ABM-FBM phase is related to the FBM-ABM phase by space parity transformation. The above described results follow through by applying the transformation $L\leftrightarrow R$.

\subsection{ABM-US and US-ABM phases}

ABM-US phase occurs when the ratio of the impurity and bulk coupling strengths corresponding to the impurities at the left and right edges is $0<(J^L_{\mathrm{imp}}/J)<\frac{4}{3}$ and  $(J^R_{\mathrm{imp}}/J)<-\frac{4}{5}$ respectively. 

In the ground state, similar to the ABM-FBM phase, the impurity at the left edge is screened because of the presence of a bound state, whereas the impurity at the right edge is unscreened. The ground state for the chain with an even number of sites contains a spinon with rapidity $\theta\rightarrow\infty$, and whose spin can be oriented in either the positive or negative $z$-direction, resulting in a four-fold degenerate ground state:

\begin{equation}
    \ket{GS}\equiv \ket{\pm 1}, \ket{0},\ket{0}'. \;\; \text{(even number of sites)}
\end{equation}

Here $\ket{0},\ket{0}'$ are symmetric and antisymmetric under the exchange of the spins of both impurities. The ground state for odd number of sites is two fold degenerate with spin $S^z=\pm\frac{1}{2}$ corresponding to the spin of the unscreened impurity. It is represented by

\begin{equation}\ket{GS}\equiv\left|{\pm\frac{1}{2}}\right\rangle.\;\; \text{(odd number of sites )}\end{equation}

Similar to the ABM-FBM phase, the impurity at the left edge can be unscreened by removing the bound state. This requires one to remove or add the spinon with rapidity $\theta\rightarrow\infty$ in the ground state for the chain with an even or odd number of sites, respectively, which costs energy $E^L_b$. Unlike the ABM-FBM phase, the impurity at the right edge cannot be screened.

The US-ABM phase is related to ABM-US phase by space parity transformation. The above described results follow through by applying the transformation $L\leftrightarrow R$.

\subsection{FBM-US, US-FBM, US-US phases}

ABM-US phase occurs when the ratio of the impurity and bulk coupling strengths corresponding to the impurities at the left and right edges is $0>(J^L_{\mathrm{imp}}/J)>-\frac{4}{5}$ and  $(J^R_{\mathrm{imp}}/J)<-\frac{4}{5}$ respectively. 

In the ground state, both impurities are unscreened. The ground state for the chain with even number of sites is four fold degenerate and is represented by

\begin{equation}
    \ket{GS}\equiv \ket{\pm 1}, \ket{0},\ket{0}', \;\; \text{(even number of sites)}
\end{equation}

where $\ket{0},\ket{0}'$ are symmetric and antisymmetric under the exchange of spins of both impurities. The ground state for the chain with odd number of sites has a spinon with rapidity $\theta\rightarrow\infty$, and whose spin is oriented either in the positive or negative $z$-direction, resulting in an eight-fold fold degenerate ground state corresponding to the spin multiplets of the highest weight states:

\begin{equation}\ket{GS}\equiv\left|{\frac{3}{2}}\right\rangle,\left|{\frac{1}{2}}\right\rangle,\left|{\frac{1}{2}}\right\rangle'.\;\; \text{(odd number of sites )}\end{equation}

Here, $\ket{\frac{1}{2}},\ket{\frac{1}{2}}'$ are symmetric and antisymmetric under the exchange of the spins of the unscreened impurities.

The impurity at the left edge can be screened by adding or removing the spinon with rapidity $\theta\rightarrow\infty$ for the chain with even and odd sites respectively, which costs energy $E^{L}_b$. 

The US-ABM phase is related to ABM-US phase by space parity transformation. The above described results follow through by applying the transformation $L\leftrightarrow R$.

The US-US phase occurs when the ratio of the impurity and the bulk coupling strengths corresponding to both impurities is $(J^i_{\mathrm{imp}}/J)<-\frac{4}{5}, i=\{L,R\}$.  The ground state is similar to the FBM-US or US-FBM phases described above, but unlike these phases, neither the left nor the right impurities can be screened. 

\section{Discussion and Outlook} 
\label{conc}

We considered the spin-$\frac{1}{2}$ Heisenberg chain with boundary impurities and analyzed it analytically using \textit{Bethe Ansatz} and computed impurity magnetization numerically using DMRG.  We found that the system exhibits different phases depending on the ratio of the boundary and bulk coupling strengths. 

In the case of one impurity interacting antiferromagnetically, there exists two phases, namely the Kondo phase and the ABM phase. The Kondo phase occurs when the ratio of the impurity and the bulk coupling strengths $(J_{imp}/J)<\frac{4}{3}$. The impurity is screened in the ground state due to the Kondo effect associated with the emergence of a strong coupling scale called the Kondo temperature $T_K$. The ratio of the impurity and the bulk density of states $R(E)$ takes a lorentzian type form when $(J_{imp}/J)$ is sufficiently small. As $(J_{imp}/J)$ is increased, $R(E)$ smoothly takes the form of a constant function as $(J_{imp}/J)$ approaches the value of $1$, where the impurity becomes a part of the bulk. As $(J_{imp}/J)$ is increased further, $R(E)$ asymptotically takes the form of a delta function peaked at $E=2\pi J$, as $(J_{imp}/J)$ approaches the value of $\frac{4}{3}$. This signifies the breakdown of the many body Kondo effect at $(J_{imp}/J)=\frac{4}{3}$, where $T_K=T_0$, which is also the maximum energy of a single spinon. Eventually as $(J_{imp}/J)$ is increased, one enters the ABM phase. The ground state in this phase contains an exponentially localized bound mode which screens the impurity. Depending on the parity of the number of sites of the chain, this mode can be removed by adding or removing a spinon, and thereby the impurity can be unscreened. This process costs energy greater than $T_0$. In the Kondo phase, excitations can be built on top of the ground state and one finds that these excitations form a tower. In the ABM phase, excitations in the bulk can be built on top of the ground state in which the impurity is screened and also on top of the state in which the impurity is unscreened, and hence one obtains two towers of excited states labelled by the total spin of the impurity, which is $S^z=0, S^z=\pm \frac{1}{2}$ respectively.

Similarly, when the impurity  interacts ferromagnetically, the system exhibits two phases: the FBM phase and the US phase, corresponding to the absolute value of the impurity and the bulk coupling strengths $|J_{imp}/J|$ being greater and lesser than $\frac{4}{5}$ respectively. The impurity is unscreened in the ground state corresponding to both the FBM and the US phases, but unlike the US phase, the impurity can be screened in the FBM phase by adding an exponentially localized bound mode by simultaneously adding or removing a spinon, depending on the parity of the number of sites of the chain. In the FBM phase, excitations can be built in the bulk on top of the ground state in which the impurity is unscreened. Similarly, excitations can also be built on top of the state in which the impurity is screened, and hence one obtains two towers of excited states labelled by the spin of the impurity, which is $S^z=\pm\frac{1}{2}, 0$ in the unscreened and the screened states respectively. In the unscreened phase, there exists a single tower corresponding to the ground state in which the impurity is unscreened. Since the Hilbert space in the ABM and the FBM phases consists of a certain number of towers, it is said to undergo ``Hilbert space fragmentation".

As one moves across the phase transition between the Kondo and the ABM phases, the ground state undergoes a characteristic change where the many body screening of the impurity turns into a single particle effect. Simultaneously, a reorganization of the Hilbert space occurs, resulting in a change in the number of towers from one to two. This phenomenon is named the `boundary eigenstate phase transition', and is also found in certain one-dimensional topological superconductors \cite{pasnoorithesis} and the spin-$\frac{1}{2}$ Heisenberg chain with boundary magnetic fields \cite{pasnoori2023boundary}. We  showed that the boundary phase transitions manifest themselves in the behavior of local observables such as  the local impurity density of states and local impurity magnetization, which  behave differently in the Kondo and in the antiferromagnetic bound mode phases.

When two impurities are considered,  they  are independent in the thermodynamic limit (up to $1/L$ corrections), and  each impurity can be in any of the four phases corresponding to the one-impurity case. Thus, the model with two boundary impurities  exhibits a total of sixteen possible phases. However, in a finite size system, the two impurities can influence each other. Analyzing the entanglement properties and dynamics across different phases can probe the novel effects arising from the mutual effects of the entanglement.  Furthermore the presence of the exponentially localized bound mode affects the dynamics of the system, whose study  could provide a comprehensive insight into the impact brought about by the eigenstate phase  transitions. We will turn  to this subject in an upcoming publication.

\section*{Acknowledgement}
We thank David Rogerson and Yicheng Tang for helpful discussions.  J.H.P. is partially supported by NSF Career Grant No.~DMR- 1941569 and the Alfred P.~Sloan Foundation through a Sloan Research Fellowship.

\bibliography{ref}

\begin{appendix}
\label{appendix}
\begin{widetext}

\section{Hamiltonian and \textit{Bethe Ansatz equations}}\label{sec: hamder}
Consider the rational six-vertex matrix $R$ 
\begin{equation}
    R_{i,j}(\lambda)=\lambda {I}_{ij}+P_{ij}
\end{equation}
where $I_{ab,cd}=\delta_{ab}\delta_{cd}$ is the identity matrix and $P_{ab,cd}=\delta_{ad}
\delta_{cb}=$ is the permutation operator. The $R$-matrix is a solution of the Yang-Baxter equation
\begin{equation}
    R_{12}(\lambda-\lambda')R_{13}(\lambda)R_{23}(\lambda')=R_{23}(\lambda')R_{13}(\lambda)R_{12}(\lambda-\lambda').
\end{equation}
Notice that, the Yang-Baxter equation remains satisfied if we shift $\lambda_i\to\lambda_i-\theta_i$ where $\theta_i$ are arbitrary inhomogeneous parameters.
and introduce two transfer matrices $T_0(\lambda)$ and $\hat T_0(\lambda)$ 
\begin{align}
T_0(\lambda)&=R_{0,L}(\lambda-b-\theta_L)R_{0,N}(\lambda-\theta_N)R_{0,N-1}(\lambda-\theta_{N-1})\cdots R_{0,2}(\lambda-\theta_2)R_{0,1}(\lambda-\theta_1)R_{0,R}(\lambda-d-\theta_R)\nonumber\\
\hat T_0(\lambda)&=R_{0,R}(\lambda+d+\theta_R)R_{0,1}(\lambda+\theta_1)R_{0,2}(\lambda+\theta_2)\cdots R_{0,N-1}(\lambda+\theta_{N-1})L_{0,N}(\lambda-\theta_N)R_{0,L}(\lambda+b+\theta_L)\nonumber
\end{align}
Now, we define the monodromy matrix
\begin{equation}
    \Xi(\lambda)=T_0(\lambda)\hat T(\lambda)
\end{equation}
The trace of the monodromy matrix over the auxiliary space is defined as the double row transfer matrix
\begin{equation}
    t(\lambda)=\operatorname{tr}_0\Xi(\lambda)
    \label{tmat}
\end{equation}
It is quite easy to see that the transfer matrix forms a one-parameter family of commuting operators \textit{i.e.}
\begin{equation}
    [t(\lambda),t(\rho)]=0.
\end{equation}
The Hamiltonian is related to the transfer matrix as
\begin{align}
    H&=J \frac{\mathrm{d}}{\mathrm{d}\lambda}\log\operatorname{tr}_0\Xi(\lambda)\big\vert_{\lambda\to0, \{\theta_i\}\to 0}-NJ-\frac{J}{1-b^2}-\frac{J}{1-d^2}\nonumber\\
    &= H=J\sum_{j=1}^{N-1}\vec\sigma_j\cdot\vec\sigma_{j+1}+\frac{J}{1-b^2}\vec\sigma_1\cdot\vec\sigma_L+\frac{J}{1-d^2}\Vec\sigma_N\cdot\vec\sigma_R
    \label{ham}
\end{align}
This shows that the Hamiltonian is integrable. Upon identifying
\begin{equation}
    J_L=\frac{J}{1-b^2} \quad \text{ and }\quad  J_R=\frac{J}{1-d^2} \label{vardef}
\end{equation}
we get the Hamiltonian Eq.\eqref{ham2imp}.

The eigen value $\Lambda(\lambda)$ of the transfer matrix $t(\lambda)$ satisfy Baxter's $T-Q$ relation \cite{baxter1972partition}
\begin{equation}
        \Lambda(\lambda)=\frac{(\lambda+1)^{2N+1}((\lambda+1)^2-b^2)((\lambda+1)^2-d^2)}{2\lambda+1}\frac{Q(\lambda-1)}{Q(\lambda)}+\frac{\left(\lambda ^2-b^2\right) \left(\lambda ^2-d^2\right) \lambda ^{2 N+1}}{2 \lambda +1}\frac{Q(\lambda+1)}{Q(\lambda)}
        \label{eval}
\end{equation}

where the Q-function is given by
\begin{equation}
Q(\lambda)=\prod_{\ell=1}^M (\lambda-\lambda_\ell	)(\lambda+\lambda_\ell+1)
\end{equation}

Regularity of the T-Q equation gives the BAEs
\begin{equation}
 \left(\frac{\lambda_j +1}{\lambda_j }\right)^{2 N+1}\frac{\lambda_j +b+1}{\lambda_j +b}\frac{\lambda_j -b+1}{\lambda_j -b}\frac{\lambda_j +d+1}{\lambda_j +d}\frac{\lambda_j -d+1}{\lambda_j -d}=-\prod_{\ell=1}^M\frac{(\lambda_j-\lambda_\ell+1)(\lambda_j+\lambda_\ell+2)}{(\lambda_j-\lambda_\ell-1)(\lambda_j+\lambda_\ell)}
\end{equation}
where $N$ is the total number of bulk spins and hence there are total $N+2$ spins in the systems including the two impurity spins. 

Changing the variable $\lambda_j=i\mu_j-\frac{1}{2}$, we rewrite the above equation as
\begin{equation}
    \left(\frac{\mu _j-\frac{i}{2}}{\mu _j+\frac{i}{2}}\right)^{2 N}\frac{\mu _j-i \left(\frac{1}{2}-b\right)}{\mu _j+i \left(\frac{1}{2}-b\right)}\frac{\mu _j-i \left(\frac{1}{2}+b\right)}{\mu _j+i \left(\frac{1}{2}+b\right)}\frac{\mu _j-i \left(\frac{1}{2}-d\right)}{\mu _j+i \left(\frac{1}{2}-d\right)}\frac{\mu _j-i \left(\frac{1}{2}+d\right)}{\mu _j+i \left(\frac{1}{2}+d\right)}=\prod_{j \neq \ell=1}^{M}\left(\frac{\mu_{j}-\mu_{\ell}-i}{\mu_{j}-\mu_{\ell}+i}\right)\left(\frac{\mu_{j}+\mu_{\ell}-i}{\mu_{j}+\mu_{\ell}+i}\right)
    \label{bae}
\end{equation}
Or, in the logarithmic form
\begin{equation}
\begin{aligned}
(2 N+1) \tan ^{-1}\left(2 \mu_j\right) & +\tan ^{-1}\left(\frac{\mu_j}{\frac{1}{2}-b}\right)+\tan ^{-1}\left(\frac{\mu_j}{\frac{1}{2}+b}\right)+\tan ^{-1}\left(\frac{\mu_j}{\frac{1}{2}-d}\right)+\tan ^{-1}\left(\frac{\mu_j}{\frac{1}{2}+d}\right) \\
& =\sum_{\ell=1}^M\left[\tan ^{-1}\left(\mu_j-\mu_{\ell}\right)+\tan ^{-1}\left(\mu_j+\mu_{\ell}\right)\right]+\pi I_j .
\end{aligned} 
\end{equation}
The density of the solution can be obtained from the integral equation which is derived by taking derivative of the above equation and substituting $\frac{\mathrm{d}}{\mathrm{d}\mu}I_j=2\rho(\mu)$. The itegral equation is of the form

\begin{equation}
    \begin{aligned}
   2\rho(\mu)&=(2N+1)a_{\frac12}(\mu)+a_{\frac{1}{2}-b}(\mu)+a_{\frac{1}{2}+b}(\mu)+a_{\frac{1}{2}-d}(\mu)+a_{\frac{1}{2}+d}(\mu)-\sum_{\upsilon=\pm}\int_{-\infty}^\infty\mathrm{d}\mu~ a_1(\mu+\upsilon\mu')\rho(\mu')-\delta(\mu)
    \end{aligned}
    \label{intbae}
\end{equation}
where
\begin{equation}
    a_\gamma(\mu)=\frac{1}{\pi}\frac{\gamma}{\mu^2+\gamma^2}
\end{equation}
where $\delta(0)$ is added to remove the $\mu_j=0$ solution which results in a vanishing wavefunction.

From Eq.\eqref{ham} and Eq.\eqref{eval}, the energy eigenvalues are obtained
\begin{equation}
    E=-\sum _{j=1}^M \frac{2J}{\mu _j^2+\frac{1}{4}}+J (N-1)+\frac{J}{1-b^2}+\frac{J}{1-d^2}
    \label{engeng}
\end{equation}

\section{Detailed solution of the\textit{Bethe Ansatz equations}}
\label{sec: DetailSol}
Here we provide the complete solution of the \textit{Bethe Ansatz equations} Eq.\eqref{bae} in various phases. The variables $b$ and $d$ introduced in Eq.\eqref{vardef} are more natural for \textit{Bethe Ansatz}. Thus, we will use these variables instead of the impurity couplings in this section. We will label the states by their spin and how they are created from the all real roots state. For example, the state $\ket{0}$ represents a a state given by all real roots solution and it has spin 0 and $\ket{\pm 1}_{bd\theta}$ represents a doubly degenerate state with spin $\pm 1$ which is obtained by adding a hole, left boundary string solution $\mu_b$ and right boundary string solution $\mu_d$ to the state given by all real root distribution.
\subsection{Kondo-Kondo phase}
When both $b$ and $d$ are purely imaginary or between $0$ and $\frac{1}{2}$, we have Kondo phases on the both edges. The solution depends on the parity of the total number of bulk sites $N$.
\subsubsection{Even number of sites}
Solving Eq.\eqref{intbae}, we obtain the solution density
\begin{equation}
   \tilde\rho_{\ket{0}}(\omega)= \frac{1}{4} \text{sech}\left(\frac{| \omega | }{2}\right) \left(2 \cosh (b | \omega | )+2 \cosh (d | \omega | )-e^{\frac{| \omega | }{2}}+2 N+1\right)
   \label{soldenskk}
\end{equation}
The total number of Bethe roots is $M=\int\rho_{\ket 0}(\mu)\mathrm{d}\mu = \tilde\rho_{\ket 0}(0)=\frac{N+2}{2}$.Thus, the ground-state magnetization
\begin{equation}
    S^z_{\ket 0}=\frac{N+2}{2}-M_{\ket 0}=0
\end{equation}
Using Eq.\eqref{engeng}, we compute the energy of the state

\begin{multline}
    E_{\ket 0}=-J\psi ^{(0)}\left(\frac{b}{2}+1\right)+J\psi ^{(0)}\left(\frac{b}{2}+\frac{1}{2}\right)-J\psi ^{(0)}\left(1-\frac{b}{2}\right)+J\psi ^{(0)}\left(\frac{1}{2}-\frac{b}{2}\right)-J\psi ^{(0)}\left(\frac{d }{2}+1\right)+J\psi ^{(0)}\left(\frac{d }{2}+\frac{1}{2}\right)\\-J\psi ^{(0)}\left(1-\frac{d }{2}\right)+J\psi ^{(0)}\left(\frac{1}{2}-\frac{d }{2}\right)
    -J((2 N+1) \log (4))+J\pi+J (N-1)+\frac{J}{1-b^2}+\frac{J}{1-d^2}
\end{multline}
There are no unique boundary excitations possible in this phase. All the excitations on top of the ground state are bulk excitations.

\subsubsection{Odd number of sites}

Since the total number of roots $\frac{N+2}{2}$ given by all real roots of the Bethe equation is not an integer for odd $N$, we need to add a hole and solve
\begin{equation}
2\rho(\mu)= (2N+1)a_{\frac12}(\mu)+a_{\frac{1}{2}-b}(\mu)+a_{\frac{1}{2}+b}(\mu)+a_{\frac{1}{2}-d}(\mu)+a_{\frac{1}{2}+d}(\mu)-\sum_{\upsilon=\pm}\int_{-\infty}^\infty\mathrm{d}\mu~ a_1(\mu+\upsilon\mu')\rho_{\left|-\frac12\right\rangle}(\mu')-\delta(\mu)-\delta(\mu-\theta)-\delta(\mu+\theta)
\nonumber
\end{equation}
such that the solution density becomes
\begin{equation}
 \tilde\rho_{\ket{\frac12}_\theta}(\omega)=   \frac{1}{4} \text{sech}\left(\frac{| \omega | }{2}\right) \left(2 \cosh (b | \omega | )+2 \cosh (d | \omega | )-e^{\frac{| \omega | }{2}} (2 \cos (\theta  \omega )+1)+2 N+1\right)
\end{equation}
Now, the total number of roots is $M=\tilde\rho_{\ket{\frac12}_\theta}(0)=\frac{N+1}{2}$ which is an integer when $N$ is odd. Thus, the ground state magnetization is
\begin{equation}
    S^z_{\ket{\frac12}_\theta}=\frac{N+2}{2}-M_{\ket{\frac{1}{2}}_\theta}=\frac{1}{2}
\end{equation}
Because of the $SU(2)$ symmetry there is another state with spin $-\frac{1}{2}$ degenerate to this Using Eq.\eqref{engeng}, the energy of these states is
\begin{equation}
E_{\ket{\pm\frac12}_\theta}=E_{\ket{0}}+\frac{2J\pi}{\cosh(\pi\theta)} 
\end{equation}
All the excitations on top of this doubly degenerate ground state are bulk excitations.

\subsubsection{Spinon density of state}
\label{DOS}
From Eq.\eqref{soldenskk}, we compute the 
density of state contribution due to the bulk
\begin{equation}
    \rho_{\mathrm{dos}}^{\mathrm{bulk}}(E) = \left| \frac{\rho_{\ket{0}^{\mathrm{bulk}}}(\mu)}{E'(\mu)} \right| = \frac{\frac{1}{2} N \text{sech}(\pi \mu)}{2 \pi^2 J \tanh (\pi \mu) \text{sech}(\pi \mu)} = \frac{N \coth (\pi \mu)}{4 \pi^2 J} = \frac{N}{2 \pi \sqrt{4 \pi^2 J^2 - E^2}}
\end{equation}
where we used the spinon energy to write $\mu(E)=\frac{1}{\pi }\cosh ^{-1}\left(\frac{2 \pi  J}{E}\right)$.
The impurity part in Kondo phase gives
\begin{align}
  { \rho_{\mathrm{dos}}}^{\mathrm{imp}}(E) &=\frac{4 \pi  J^2 \cos (\pi  b)}{\sqrt{4 \pi ^2 J^2-E^2} \left(E^2 \cos (2 \pi  b)-E^2+8 \pi ^2 J^2\right)}
\end{align}

Let us now consider the ratio
\begin{align}
   R(E)= \frac{N}{2}\frac{{ \rho_{\mathrm{dos}}}^{\mathrm{imp}}(E)}{{ \rho_{\mathrm{dos}}}^{\mathrm{bulk}}(E)}&=\frac{4 \pi ^2 J^2 \cos (\pi  b)}{E^2 \cos (2 \pi  b)-E^2+8 \pi ^2 J^2}
\end{align}
from which we get Eq.\eqref{redef} in the main text by writting in terms of the ratio of the couplings. 

Notice that the spinon energy is bounded $0<E<2\pi J$. The function $R(E)$ is maximum at $E=0$ and minimum at $E=2\pi J$ for purely imaginary $b$ but it is mximum at $E=2\pi J$ and minimum at $E=0$ when $0< b<\frac{1}{2}$. We compute the Kondo temperature as the energy scale  at which the integrated density of state is half of the total number of state
 
\begin{equation}
    \int_0^{T_K} d E \rho_{i m p}(E)=\frac{1}{2} \int_0^{T_0} d E \rho_{i m p}(E)
\end{equation}

where $T_0=2 \pi J$ is the bandwidth of the single spinon branch. This yields for $T_K$
\begin{equation}
 T_K=\frac{T_0}{\sqrt{1+\cosh ^2(\pi \beta)}}   
\end{equation}

\subsection{Kondo-Bound mode phase}
\label{sec: Kondo-Bound mode}
When $\frac{1}{2}<b<1$ and $d$ is purely imaginary or $0<d<\frac{1}{2}$ or vice versa, we have the Kondo effect at one end of the chain and the formation of the bound mode at another end. Here we solve the equation for $\frac{1}{2}<d<1$ and $0<b<\frac{1}{2}$ and all other cases can be obtained by analytic continuation and left/right symmetry (\textit{i.e.} exchanging $b$ and $d$). 
\subsubsection{Even number of sites}
The Bethe equation
\begin{equation}
    \left(\frac{\mu _j-\frac{i}{2}}{\mu _j+\frac{i}{2}}\right)^{2 N}\frac{\mu _j-i \left(\frac{1}{2}-d\right)}{\mu _j+i \left(\frac{1}{2}-d\right)}\frac{\mu _j-i \left(\frac{1}{2}+d\right)}{\mu _j+i \left(\frac{1}{2}+d\right)}\frac{\mu _j+i \left(b-\frac{1}{2}\right)}{\mu _j-i \left(b-\frac{1}{2}\right)}\frac{\mu _j-i \left(\frac{1}{2}+b\right)}{\mu _j+i \left(\frac{1}{2}+b\right)}=\prod_{j \neq \ell=1}^{M}\left(\frac{\mu_{j}-\mu_{\ell}-i}{\mu_{j}-\mu_{\ell}+i}\right)\left(\frac{\mu_{j}+\mu_{\ell}-i}{\mu_{j}+\mu_{\ell}+i}\right)
\end{equation}
admits an additional solution 
\begin{equation}
    \mu_d=\pm i \left(\frac12-d\right)
\end{equation}

This kind of solution is responsible for boundary excitation in various integrable models with open boundary conditions \cite{kapustin1996surface,kattel2023exact,pasnoori2022rise,pasnoori2023boundary}.

As we see below, the boundary string has negative energy. Thus, we solve the root density of the Bethe equation by adding this boundary string solution. We obtain
\begin{equation}
    \tilde\rho_{\ket{0}_d}(\omega)= \frac{1}{4} \text{sech}\left(\frac{| \omega | }{2}\right) \left(-e^{(d-1) | \omega | }-e^{| \omega | -d | \omega | }+e^{-b | \omega | }+e^{b | \omega | }-e^{\frac{| \omega | }{2}}+2 N+1\right)
\end{equation}

The total number of the Bethe roots is
    $M=1+\tilde\rho_{\ket{0}_d}(0)=\frac{N+2}{2}$
which results in vanishing ground state magnetization
\begin{equation}
    S^z_{\ket{0}_d}=\frac{N+2}{2}-M_{\ket{0}_d}=0
\end{equation}

showing that both impurities are screened. The right impurity is screened by the bound mode formed at the impurity site whereas the left impurity is screened by the multiparticle Kondo cloud. 

The energy of this state is
\begin{equation}
    E_{\ket{0}_d}=E_{\ket{\pm \frac{1}{2}}}+E_d
\end{equation}
where
\begin{multline}
   E_{\ket{\pm \frac{1}{2}}}= -\frac{2J}{d}-2 J \psi ^{(0)}\left(\frac{d}{2}\right)+2 J \psi ^{(0)}\left(\frac{d}{2}+\frac{1}{2}\right)+J (N-1)+\frac{J}{1-b^2}+\frac{J}{1-d^2}\\-J\psi ^{(0)}\left(\frac{b }{2}+1\right)+J\psi ^{(0)}\left(\frac{b }{2}+\frac{1}{2}\right)-J\psi ^{(0)}\left(1-\frac{b }{2}\right)+J\psi ^{(0)}\left(\frac{1}{2}-\frac{b }{2}\right)-(2 N+1) J\log (4)+J\pi\\
\end{multline}
is the energy of all real roots and
\begin{equation}
    E_d=-2\pi J \csc(\pi d)
\end{equation}
is the energy of the bound mode. 

\begin{figure}[H]
\centering
\begin{tikzpicture}
  \begin{axis}[
    axis lines=center,
    xlabel={$d$},
    ylabel={$\frac{E_d}{J}=-\frac{2\pi}{\sin(\pi d)}$},
    domain=0.5:0.999999,
    xmin=0.5, xmax=0.999999,
    ymin=-200, ymax=50,
    samples=400,
    smooth,
    thick,
    ]
    \addplot [color=red, thick] {-(2*pi)/(sin(deg(pi*x)))};
  \end{axis}
\end{tikzpicture}
\caption{Boundary string energy}
\end{figure}
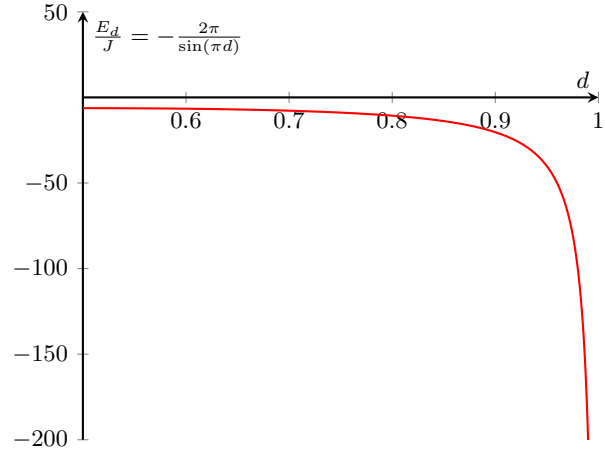

Removing the boundary string and adding a hole to all real root distribution, we obtain a state described by the root density of the form
\begin{equation}
   \tilde\rho_{\ket{1}_\theta}(\omega)=  \frac{1}{4} \text{sech}\left(\frac{| \omega | }{2}\right) \left(\left(e^{| \omega | }-1\right) \left(-e^{-d | \omega | }\right)+e^{-b | \omega | }+e^{b | \omega | }-2 e^{\frac{| \omega | }{2}} \cos (\theta  \omega )-e^{\frac{| \omega | }{2}}+2 N+1\right)
\end{equation}

The spin of the state is $S^z_{\ket{1}_\theta}=1$. Here the right impurity is unscreened and forms triplet pairing with the propagating hole. 

The energy of this state is
\begin{equation}
    E_{\ket{1}_\theta}= E_{\ket{\pm \frac{1}{2}}}+E_\theta
    \label{tripkbm}
\end{equation}
where $E_\theta=\frac{2\pi}{\cosh(\pi\theta)J}$ is the energy of a hole, and $E_d =-\frac{2\pi J}{\sin(\pi d)}$ is the energy of the bound mode. Notice that due to the $SU(2)$ symmetry, there also exist $\ket{-1}_\theta$ and $\ket{0}_\theta$ states with spins $S^z=-1$ and $S^z=0$, respectively, having the same energy $E_{\ket{1}_\theta}$.

After adding the boundary string solution $\mu_d$, the resultant Bethe equation admits higher order boundary string solution
\begin{equation}
 \mu_{dh}=   \pm i \left(\frac{3}{2}-d\right)
\end{equation}
Adding a hole and both string solutons $\mu_d$ and $\mu_{dh}$, we obtain a strate describe by the root density
\begin{equation}
  \tilde\rho_{\ket{0}_{d,dh,\theta}}(\omega)  =\frac{1}{4} \text{sech}\left(\frac{| \omega | }{2}\right) \left(-e^{(d-2) | \omega | }-2 e^{-(d-1) | \omega | }-e^{(d-1) | \omega | }+e^{-b | \omega | }+e^{b | \omega | }-2 e^{\frac{| \omega | }{2}} \cos (\theta  \omega )-e^{\frac{| \omega | }{2}}+2 N+1\right)
\end{equation}
with spin $S^z=0$ and energy
\begin{equation}
E_{\ket{0}_{d,dh,\theta}}=E_{\ket{1}_\theta}
\label{singkbm}
\end{equation}

The energy of the $\mu_{dh}$ solution is exactly negative of the $\mu_d$ solution. Thus, adding both solutions results in a state where right impurity is unscreened and combines with the propagating spinon to form singlet pairing. The energies of the singlet and triplets are the same in thermodynamics limit. 

Apart from these boundary excitations where an unscreened impurity and a propagating spinon form $S^z=0,\pm 1$, there are other excitations which can be constructed by adding an even number of spinons, bulk strings, quartets, and other higher-order boundary strings. 

\subsubsection{Odd number of sites}
In the case of an odd number of sites, the ground state is constructed by adding the boundary string solution $\mu_d$ and a propagating hole on top of the real solution. 

This forms a doubly degenerate ground state with spin
\begin{equation}
    S^z_{\ket{\pm \frac{1}{2}}_{d,\theta}}=\pm \frac{1}{2}
\end{equation}
with energy
\begin{equation}
    E_{\ket{\pm \frac{1}{2}}_{d,\theta}}=E_{\ket{0}_d}+E_\theta
    \label{engbtodd}
\end{equation}
Boundary excitations can be constructed by removing both the boundary string and holes such that a doubly degenerate state with spin
\begin{equation}
    S^z_{\ket{\pm \frac{1}{2}}}=\pm \frac{1}{2}
\end{equation}
can be constructed where one of the impurities is unscreened. The energy of this state is $E_{\ket{\pm \frac{1}{2}}}$.
\subsection{Kondo-Ferromagnetic bound mode phase}
When $b$ is purely imaginary or $0<b<\frac{1}{2}$ and $1<d<\frac{3}{2}$ (or vice versa), there exists multiparticle screening of impurity at one edge, whereas the other impurity can only be screened at high energy. 

The main difference between this phase and the Kondo-bound mode phase (see Section \ref{sec: Kondo-Bound mode}) is that the boundary string has a positive energy. 

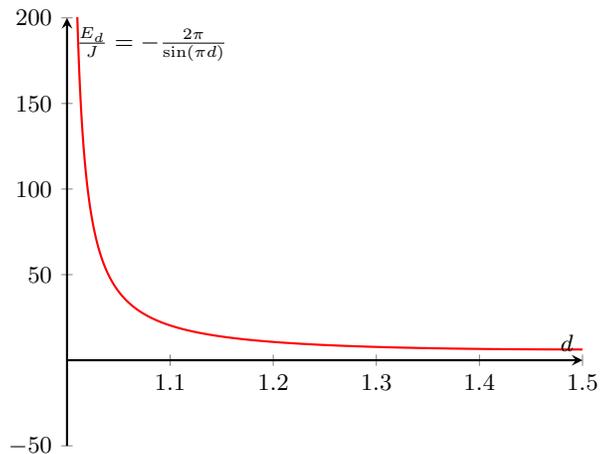
\begin{figure}[H]
\centering
\begin{tikzpicture}
  \begin{axis}[
    axis lines=center,
    xlabel={$d$},
    ylabel={$\frac{E_d}{J}=-\frac{2\pi}{\sin(\pi d)}$},
    domain=1:1.5,
    xmin=1, xmax=1.5,
    ymin=-50, ymax=200,
    samples=400,
    smooth,
    thick,
    ]
    \addplot [color=red, thick] {-(2*pi)/(sin(deg(pi*x)))};
  \end{axis}
\end{tikzpicture}
\caption{Boundary string energy}
\end{figure}
Thus, the ground state contains one unscreened impurity spin and another impurity screened by Kondo cloud. However, there exists a high energy excited state where the a bound mode is formed at the site of the impurity and hence the impurity gets screened. Let us consider the cases of even and odd number of sites separately. 

\subsubsection{Even number of sites}
For an even number of sites, there are four-fold degenerate states $\ket{\pm 1}_\theta$ with energy given by Eq. \eqref{tripkbm}, and $\ket{0}_{d,dh,\theta}$ with energy given by Eq. \eqref{singkbm}. In the latter, left impurity is screened by multiparticle Kondo screening, while the right impurity is unscreened and it forms singlet or triplet pairs with a propagating ground state, resulting in a four-fold degenerate ground state.

We can remove the propagating hole from the ground state and add the boundary string solution $\mu_d$ to form a high energy state with energy given by $E_{\ket{0}_d}$ where both the impurities are screened.

\subsubsection{Odd number of sites}
The state $\ket{\pm \frac{1}{2}}$ is described by all real roots and has energy $E_{\ket{\pm \frac{1}{2}}}$. This state is the doubly degenerate ground state, where the left impurity is unscreened, and the right impurity is screened by a multiparticle Kondo cloud.

By adding a hole and the boundary string, we can construct a high-energy state $\ket{\pm \frac{1}{2}}_{b,\theta}$ with energy given by Eq. \eqref{engbtodd}. In this state, the right impurity is screened by multiparticle screening, and the left impurity is screened by the bound mode formed at the impurity site. This state exhibits a propagating spinon, resulting in its doubly degenerate nature.

\subsection{Kondo-Unscreened Phase}
Here we consider the case where $b$ is purely imaginary or in the range $0<b<\frac{1}{2}$ and $d>\frac{3}{2}$. In this phase, the boundary string solution
\begin{equation}
    \mu_d=\pm i\left(\frac{1}{2}-d\right)
\end{equation}
becomes a wide boundary \cite{destri1982analysis} and thus it has zero energy. This wide boundary string with vanishing energy cannot screen the left impurity at any energy scale. 
\subsubsection{Even number of sites}
The ground state in this phase is exactly like in Kondo-Ferromagnetic bound mode phase \textit{i.e} the ground state contains left impurity whose spin is screened by multiparticle Kondo cloud while the right impurity is unscreened and it forms singlet or triplet pairing with a propagating spinon making the ground state four-fold degenerate. The state where impurity forms triplet pair with the propagating spinon is constructed by adding a hole to all real root solution which gives a state with spin $S^z=\pm 1,0$ and and energy
\begin{equation}
    E_{\ket{\pm 1, 0}_\theta}=E_{\ket{1}_\theta}.
\end{equation}
The state where the unscreened impurity and the spinon forms singlet pair is constructed by adding a hole and the wide-boundary string $\mu_b$ to all real roots solution which gives a state with spin $S^z=0$ and energy
\begin{equation}
    E_{\ket{0}_{\theta,d}}=E_{\ket{1}_\theta}
\end{equation}
All other excitation can be constructed by adding even number of spinons, bulk strings and quartets. 

\subsubsection{Odd number of sites}
When the total number of sites is odd, the ground state is made up of all real roots solution with spin $S^z=\pm \frac{1}{2}$ and energy $E_{\ket{\pm \frac{1}{2}}}$. 

All excitations on top of this doubly degenerate ground state can be constructed by adding the even number of spinons, bulk strings, quartets, and the wide-boundary string or higher boundary string with even number of spinons. 

\subsection{Other phases}
Since, we consider all possible distinct phases that can occur in one end in the previous sections, now we can briefly discuss the solution of \textit{Bethe Ansatz equations} \eqref{bae} in remaining phases. We will only consider the case where $N$ is even for the remaining of the phases.
\subsubsection{Bound mode- Bound mode phase}
When both $b$ and $d$ take values between $\frac{1}{2}$ and 1, the ground state is made up of all real roots and two boundary string solutions $\mu_b=\pm i\left(\frac{1}{2}-b\right)$ and $\mu_d=\pm i\left(\frac{1}{2}-d\right)$. The state is described by the solution density of the form
\begin{equation}
  \tilde\rho_{\ket{0}_{b,d}}(\omega)   =\frac{1}{4} \text{sech}\left(\frac{| \omega | }{2}\right) \left(-2 \cosh ((b-1) | \omega | )-2 \cosh ((d-1) | \omega | )-e^{\frac{| \omega | }{2}}+2 N+1\right)
\end{equation}
The total spin of the state is 
\begin{equation}
    S^z_{\ket{0}_{b,d}} 
\end{equation}
and the total energy is
\begin{equation}
    E_{\ket{0}_{b,d}}=E_{\ket{\pm 1}}+E_b+E_d
\end{equation}
where 
\begin{multline}
   E_{\ket{\pm 1}}=-2 \pi J  \csc (\pi  d)-2 \pi J  \csc (\pi  d)-\frac{2J}{d}-\frac{2J}{b}-2 J \psi ^{(0)}\left(\frac{d}{2}\right)+2 J \psi ^{(0)}\left(\frac{d}{2}+\frac{1}{2}\right)+J (N-1)+\frac{J}{1-d^2}+\frac{J}{1-b^2}\\-\frac{2 J}{b}-2 J \psi ^{(0)}\left(\frac{b}{2}\right)+2 J \psi ^{(0)}\left(\frac{b}{2}+\frac{1}{2}\right)-(2 N+1) J\log (4)+J\pi
\end{multline}
is the energy of all real roots and
\begin{equation}
    E_b=-2\pi J\csc(\pi b) \quad\text{ and }\quad E_d=-2\pi J\csc(\pi d)
\end{equation}
are the energies of the boundary strings.  Both impurities are screened by bound modes formed at the impurity sites in the ground state. 

We can remove the boundary string $\mu_b$ and add a hole to construct a state $\ket{\pm 1}_{d,\theta}$ with spin $S^z_{\ket{\pm 1}_{d,\theta}}=\pm 1$ with energy
\begin{equation}
    E_{\ket{1}_{d,\theta}}= E_{\ket{\pm 1}}+E_\theta+E_d
\end{equation}
This is a state where left impurity is unscreened and forms triplet pairing with a propagating hole while the right impurity is screened by the bound mode. Moreover, by adding a hole, the boundary string solution $\mu_b$ and the higher order boundary string solution $\mu_{bh}$ we can construct a state where the unscreened impurity forms a singlet pair with the propagating hole. The singlet and triplets have the same energy in the thermodynamic limit. 

We could remove the boundary solution $\mu_d$ from the ground state and add a hole to construct a state with spin $S^z_{\ket{\pm 1}_{b,\theta}}=\pm 1$ with energy
\begin{equation}
E_{\ket{1}_{b,\theta}}= E_{\ket{\pm 1}}+E_\theta+E_b
\end{equation}
In this state the left impurity is screened by bound mode while the right impurity is unscreened and it forms triplet pairings with a propagating hole. By adding a hole, the boundary string $\mu_d$ and the higher order boundary string $\mu_{dh}=\pm i\left(\frac{3}{2}-d\right)$, we can create a state where the left impurity is screened and right impurity is unscreened and formed singler pairing with a propagating spinon. 

We could remove both the $\mu_d$ and $\mu_b$ boundary strings and create an excitation where both impurities are unscreened and form tripet pairings. To constructed a state where these two unscreened impurity are unscreened, we need to add higher order boundary strings. 

All other excitations can be constructed by adding even number of spinons, bulk strings, quartets, and higher order boundary strings. 

\subsubsection{Bound mode - Ferromagnetic bound mode phase}
When $\frac{1}{2}<d<1$ and $1<b<\frac{3}{2}$ or vice versa , the ground state contains one impurity screened by bound mode while the other one is unscreened and forms a singlet or triplet pairing with a spinon, thereby forming a four-fold degenerate ground state. 

Various excited states are formed by either screening the previously unscreened impurity or unscreening the previously screened impurity, or both by appropriately adding or removing boundary string, higher-order boundary string and holes.  

\subsubsection{Bound mode - Unscreened phase}
When $\frac{1}{2}<d<1$ and $b>\frac{3}{2}$ or vice versa , the ground state contains one impurity screened by the bound mode, while the other one is unscreened and forms a singlet or triplet pairing with a spinon, thereby forming a four-fold degenerate ground state. 

Because the unscreened impurity cannot be screened at any energy scale in this regime, the only possible boundary excitation involves unscreening the screened impurity. The unscreened impurity can form singlet or triplet pairings with a spinon, thereby making this excited state four-fold degenerate. 

All other excitations can be constructed by adding even number of spinons, bulk strings, quartets, and higher order boundary strings. 

\subsubsection{Ferromagnetic bound mode- Ferromagnetic bound mode phase}
When both $d$ and $b$ take values between $1$ and $\frac{3}{2}$, the ground state contains two unscreened impurities. The two unscreened impurities form a four-fold degenerate ground state. 

Boundary excitations involve screening of one or both impurities by adding appropriate boundary strings. 

All other excitations can be constructed by adding even number of spinons, bulk strings, and quartets. 

\subsubsection{Ferromagnetic bound mode - Unscreened phase}

When $1<d<\frac{3}{2}$ and $b$ take values between $1$ and $\frac{3}{2}$ or vice versa, the ground state contains two unscreened impurities. The two unscreened impurities form a four-fold degenerate ground state. 

For the case $1<d<\frac{3}{2}$ and $b>\frac{3}{2}$, the boundary excitation involves screening of the right impurity by adding a boundary string $\mu_d=\pm i \left(\frac{1}{2}-d\right)$. The left impurity cannot be screened at any energy scale.

All other excitations can be constructed by adding even number of spinons, bulk strings, and quartets.

\subsubsection{Unscreened-Unscreened phase}
When both $d$ and $b$ take values larger than $\frac{3}{2}$, the ground state contains two unscreened impurities. The two unscreened impurities form a four-fold degenerate ground state.

All other excitations can be constructed by adding even number of spinons, bulk strings, quartets, and higher order boundary strings.

\section{Summary of low lying excitation when $N$ is even}
 We introduce $K$, $BM$, $FBM$ and $US$ to represent the Kondo, bound mode, ferromagnetic bound mode and unscreened phase respectively. Moreover, the phase $K-BM$ represents the Kondo phase at the left edge and the bound mode phase at the right end. Provided the information about the $K-BM$ phase, we can extract the information about the $BM-K$ phase \textit{i.e.,} bound mode phase at the left edge and the Kondo phase at the right end by replacing $b$ and $d$ by utilizing the left/right symmetry of the model.
\begin{table}[ht]
    \centering
    \begin{tabular}{|c|c|c|c|c|}
        \hline
        \hline
        Phase & States  & Energy & Left Impurity & Right Impurity \\
        \hline
        \hline
       K-K Phase & $\ket{0}$ & $E_{\ket{0}}$ (GS) & Screened  & Screened \\
        \hline
        \hline
       K-BM Phase & $\ket{0}_d$ & $E_{\ket{0}_d}$ (GS) & Screened & Screened \\
        \cline{2-5}
                               & $\ket{\pm 1,0}_\theta$ & $E_{\ket{1}_\theta}$ & Screened & Unscreened\\
        \hline
        \hline
       K-FBM Phase & $\ket{\pm 1,0}_\theta$ & $E_{\ket{1}_\theta}$ (GS) & Screened & Unscreened\\
        \cline{2-5}
                                     & $\ket{0}_d$ & $E_{\ket{0}_d}$ & Screened & Screened \\
        \hline
        \hline
       K-US Phase & $\ket{\pm 1,0}_\theta,\ket{0}_{d,\theta}$ & $E_{\ket{1}_\theta}$ (GS) & Screened & Unscreened \\
        \hline
        \hline
        BM-BM Phase & $\ket{0}_{b,d}$ & $E_{\ket{0}_{b,d}}$ (GS) & Screened & Screened \\
        \cline{2-5}
                    & $\ket{\pm 1,0}_{d,\theta}, \ket{0}_{b,bh,d,\theta}$ & $E_{\ket{1}_{d,\theta}}$ & Unscreened & Screened \\
        \cline{2-5}
                    & $\ket{\pm 1,0}_{b,\theta}, \ket{0}_{d,dh,b,\theta}$ & $E_{\ket{1}_{b,\theta}}$  & Screened & Unscreened \\
        \cline{2-5}
                    & $\ket{\pm 1,0},\ket{0}_{b,bh}$ & $E_{\ket{\pm 1}}$ & Unscreened & Unscreened \\
        \hline
        \hline
         BM-FBM Phase & $\ket{\pm 1,0}_{b,\theta}, \ket{0}_{d,dh,b,\theta}$ & $E_{\ket{1}_{b,\theta}}$ (GS) & Screened & Unscreened \\
        \cline{2-5}
                    & $\ket{0}_{b,d}$ & $E_{\ket{0}_{b,d}}$ & Screened & Screened \\
        \cline{2-5}
                    & $\ket{\pm 1,0},\ket{0}_{b,bh}$ & $E_{\ket{\pm 1}}$ & Unscreened & Unscreened \\
        \cline{2-5}
                    & $\ket{\pm 1,0}_{d,\theta}, \ket{0}_{b,bh,d,\theta}$ & $E_{\ket{1}_{d,\theta}}$ & Unscreened & Unscreened \\
        \hline
        \hline
        FBM-FBM Phase & $\ket{\pm 1,0},\ket{0}_{b,bh}$ & $E_{\ket{\pm 1}}$ (GS) & Unscreened & Unscreened \\
        \cline{2-5}
                    & $\ket{\pm 1,0}_{d,\theta}, \ket{0}_{b,bh,d,\theta}$ & $E_{\ket{1}_{d,\theta}}$ & Unscreened & Screened \\
        \cline{2-5}
                    & $\ket{\pm 1,0}_{b,\theta}, \ket{0}_{d,dh,b,\theta}$ & $E_{\ket{1}_{b,\theta}}$  & Screened & Unscreened \\
        \cline{2-5}
                    & $\ket{0}_{b,d}$ & $E_{\ket{0}_{b,d}}$ (GS) & Screened & Screened \\
        \hline
        \hline
        BM-US Phase & $\ket{0}_{bd\theta},\ket{\pm 1,0}_{b,\theta}$ & $E_{\ket{1}_{b,\theta}}$ (GS) & Screened & Unscreened \\
        \cline{2-5}
                               & $\ket{\pm 1,0},\ket{0}_d $ & $E_{\ket{\pm 1}}$ & Unscreened & Unscreened\\
        \hline
        \hline
        FBM-US Phase  & $\ket{\pm 1,0},\ket{0}_d $ & $E_{\ket{\pm 1}} (GS)$ & Unscreened & Unscreened\\
        \cline{2-5}
                              & $\ket{0}_{bd\theta},\ket{\pm 1,0}_{b,\theta}$ & $E_{\ket{1}_{b,\theta}}$ & Screened & Unscreened \\
        \hline
        \hline
        US-US Phase & $\ket{\pm 1,0}, \ket{0}_b$  & $E_{\ket{\pm 1}}$ (GS) & Unscreened & Unscreened \\
        \hline
        \hline
    \end{tabular}
    \caption{ Energies and state of two impurities in the ground
state and the lowest energy states corresponding to each tower in all
the phases for an even number of sites.}
    \label{tab:states}
\end{table}

\end{widetext}

\end{appendix}

\end{document}